\documentclass{article}
\usepackage{lipsum}
\usepackage{multicol}
\usepackage{amsmath}
\usepackage[utf8]{inputenc}
\usepackage{graphicx}
\usepackage{epstopdf}
\usepackage{booktabs}  
\usepackage{amsmath}
\usepackage{textcomp}  
\usepackage{xcolor}
\usepackage{lipsum}    
\usepackage[Lenny]{fncychap}
\usepackage{longtable}
\usepackage{bm}
\usepackage{adjustbox}
\usepackage{mathrsfs}
\usepackage{parskip}
\usepackage[breaklinks=true]{hyperref}

\usepackage{varwidth}
\usepackage{geometry}
\begin{document}

\title{Parameter inference in a computational model of hemodynamics in pulmonary hypertension}
\date{}

\author{Amanda L. Colunga$^*$$^{1}$,
Mitchel J. Colebank$^*$$^{1,2}$, \\
REU Program$^{1}$,
Mette S. Olufsen$^{1}$ \\ \\
\small{$^{1}$Department of Mathematics, North Carolina State University, Raleigh, NC}\\
\small{$^{2}$Edwards Lifesciences Foundation Cardiovascular Innovation and Research Center, } \\ \small{and Department of Biomedical Engineering, University of California, Irvine, Irvine, CA}\\
\small{$^*$ authors contributed equally}\\ 
Correspondence: Mette S. Olufsen msolufse@ncsu.edu}

\maketitle



\begin{abstract}
Pulmonary hypertension (PH), defined by a mean pulmonary arterial pressure (mPAP) $>$ 20 mmHg, is characterized by increased pulmonary vascular resistance and decreased pulmonary arterial compliance. There are few measurable biomarkers of PH progression, but a conclusive diagnosis of the disease requires invasive right heart catheterization (RHC). Patient-specific computational models of the cardiovascular system are a potential noninvasive tool for determining additional indicators of disease severity. Using computational modeling, this study quantifies physiological parameters indicative of disease severity in nine PH patients. The model includes all four heart chambers and the pulmonary and systemic circulations. We consider two sets of calibration data: static (systolic \& diastolic values) RHC data and a combination of static and continuous, time-series waveform data. We determine a subset of identifiable parameters for model calibration using sensitivity analyses and multistart inference, and carry out uncertainty quantification post-inference. Results show that additional waveform data enables accurate calibration of the right atrial reservoir and pump function across the PH cohort. Model outcomes, including stroke work and pulmonary resistance-compliance relations, reflect typical right heart dynamics in PH phenotypes. Lastly, we show that estimated parameters agree with previous, non-modeling studies, supporting this type of analysis in translational PH research.
\end{abstract}


\begin{tabular}{cl}
\textbf{Abbreviations} &  \\
0D & Zero dimensional (time or spatial component) \\
CO & Cardiac output \\ 
CTEPH & Chronic thromboembolic pulmonary hypertension \\
iid	& Independent and identically distributed \\
MPA	& Main pulmonary artery \\
mPAP & Mean pulmonary arterial pressure \\
ODE	& Ordinary differential equation \\
PAH	& Pulmonary arterial hypertension \\
PAWP & Pulmonary arterial wedge pressure \\
PH & Pulmonary hypertension \\
PVR	& Pulmonary vascular resistance \\ 
RA	& Right atrium \\
RHC	& Right heart catheterization \\
RV	& Right ventricle \\
\end{tabular}

\section{Introduction}

Patients with a resting mean pulmonary arterial blood pressure (mPAP) greater than 20 mmHg are diagnosed with pulmonary hypertension (PH) \cite{Simonneau2019a}. This disease has no cure and, if left untreated, progresses rapidly, leading to thickening and stiffening of the pulmonary vasculature, vascular-ventricular decoupling, and right ventricular (RV) failure \cite{Fayyaz2018,Hoeper2017}. There are five main PH etiologies: pulmonary arterial hypertension (PAH, group 1), PH due to left heart disease (group 2), PH due to lung disease and/or hypoxia (group 3), chronic thromboembolic PH (CTEPH, group 4), and PH with unclear multifactorial mechanisms (group 5) \cite{Foshat2017}. Only patients in groups 1 and 4 have PH as their primary disease; in groups 2-5, PH is a comorbidity. Patients with PAH and CTEPH experience common symptoms early on, including shortness of breath, dizziness, fainting, fatigue, and swelling of the legs and abdomen \cite{Peacock2016}. Early diagnosis is difficult. Therefore patients with suspected PH undergo several tests. A definite diagnosis requires invasive pulmonary arterial blood pressure measurements through right heart cardiac catheterization (RHC) \cite{Peacock2016,Mandras2020}. PH symptoms do not appear until 1-2 years after disease onset \cite{Hoeper2017}. At this time, patients have typically undergone significant disease progression; before diagnosis limiting and reducing treatment outcomes. Understanding how cardiovascular parameters (e.g., pulmonary vascular resistance (PVR) and compliance) are modulated with the disease can assist in early detection and better therapeutic interventions. We utilize systems-level computational models with RHC data to study how model parameters and outcomes are modulated with PH.

Mathematical modeling is useful for monitoring and understanding cardiovascular disease progression. Systems-level models with multiple cardiovascular compartments have had notable success in analyzing in-vivo dynamics \cite{Kung2014,Shimizu2010,Colunga2020}. For example, Colunga et al. \cite{Colunga2020} utilized a zero-dimensional (0D) systems-level model to predict pressure-volume (PV) loops and left ventricular (LV) power to understand heart transplant recovery. Kung et al. \cite{Kung2014} used a similar model to quantify exercise capacity in Fontan patients, an essential indicator of patient survival. The study by Shimizu et al. \cite{Shimizu2010} used a 0D model to study postoperative dynamics in patients with a hypoplastic RV. Their results show that the effectiveness of ventricular repair can be predicted by RV stiffness. These studies used models to predict patient outcomes. As noted by Colunga et al. \cite{Colunga2020}, reliable results require that model parameters are identifiable given the model structure and available data. Parameters are identifiable if they influence the model output and can be uniquely determined by available data. A parameter's influence on model predictions is quantified using local \cite{Ellwein2008,Olufsen2013} and global \cite{Eck2016,Campos2020,Calvo2018} sensitivity analyses. Subset selection algorithms \cite{Olufsen2013,Miao2011} determine parameter interdependence and reduce identifiability issues. Schiavazzi et al. \cite{Schiavazzi2017} estimated cardiovascular model parameters by fitting simulations to data from single-ventricle patients with a Norwood physiology. They show that combining local and global identifiability techniques, {\it apriori}, provides unique and consistent parameter estimates given the available data. Our group \cite{Colunga2020} used similar methods to analyze data from heart-transplant patients finding that model predictions align with static RHC data measured at one point and over longitudinal patient recordings. 

These previous studies use noninvasive or static data, while others used dynamic time-series data, such as pressure waveforms, for model calibration. Marquis et al. \cite{Marquis2018} developed a compartment model of the systemic circulation. The model was calibrated by inferring five identifiable model parameters to simultaneously recorded LV pressure and volume waveforms in rats. Their results showed that estimating these parameters led to agreement between the dynamic model prediction and the waveform data. The study by Bj{\o}rdalsbakke et al. \cite{Bjordalsbakke2022} compared model sensitivity using static or dynamic outputs from a systemic circulation model. They found that time-averaged global sensitivities of aortic pressure were less influential to systemic resistance than static systolic and diastolic pressure outputs. Gerringer et al. \cite{Gerringer2018} used three- and four-element Windkessel models to predict main pulmonary artery (MPA) pressure waveforms in control and PAH mice. The study matched model simulations to dynamic MPA data, showing good agreement with the data. However, the authors did not consider a closed-loop model. These studies demonstrate the importance of employing sensitivity analyses and parameter reduction but do not discuss what data, static and/or dynamic, are informative for parameter inference. Most clinical protocols only utilize static data in electronic health records. Though static measurements are extracted from waveform data, storing patient static and dynamic pressure adds complexity to data storage. However, PH time-series pressure data may reveal important markers of disease severity.

The objective of this study is two-fold: we 1) investigate if systems-level model calibration is improved by adding dynamic RHC data and 2) investigate if patient-specific cardiovascular parameters are consistent with the physiological understanding of PH. To do so, we study the impact of model parameters on hemodynamic predictions using local and global sensitivity analyses. To quantify the benefits of adding waveform data in parameter inference, we consider two residual vectors: comparing model predictions to static data (systolic, diastolic, and mean pressures and cardiac output (CO)) and using a combination of static and dynamic data (RHC time-series waveforms). By integrating mathematical modeling, patient-specific data, and physiological intuition, we categorize each patient's functional state, including right atrial (RA), RV, and pulmonary artery (PA) temporal dynamics. In addition, we run simulations with estimated parameters to calculate patient-specific physiological biomarkers, including PV loops and other markers of PH severity.

\section{Methods}
\subsection{Ethics and approval}
Patient-specific data are obtained from two hospitals, adhering to their respective institutional review board guidelines. Deidentified RHC patient data are obtained from the Scottish Pulmonary Vascular Unit at Golden Jubilee National Hospital, Glasgow, UK, and from the Center for Pulmonary Vascular Disease at Duke University Medical Center, Durham, NC.

\subsection{Blood pressure data}
This study utilizes clinically deidentified RHC data from nine patients with confirmed PH: five with PAH and four with CTEPH. Three CTEPH and three PAH datasets are from Duke University, and one CTEPH and two PAH datasets are from the Scottish Pulmonary Vascular Unit. Static data include height, weight, sex, age, heart rate, systolic, diastolic, and mean systemic blood pressure measured by cuff pressure. The patients underwent RHC, during which a catheter was advanced from the RA to the RV and MPA. Dynamic pressure waveforms are recorded in each compartment. The pulmonary arterial wedge pressure (PAWP, mmHg), an estimate of left atrial pressure, is also recorded. CO (L/min) is measured during RHC by thermodilution. All pressure readings are obtained over 7-8 heartbeats. Demographics are provided in table \ref{tab:Demographic}. 
\begin{table}[!htp] 
   \caption{Patient demographics; group 1: pulmonary arterial hypertension (PAH); group 4: chronic thromboembolic pulmonary hypertension (CTEPH).} \label{tab:Demographic}
    \begin{center}
    \begin{tabular}{c|cclcrc} \midrule
    \textbf{Patient} & \textbf{PH} & \textbf{Age} & \textbf{Sex} & \textbf{Height (cm)} & \textbf{Weight (kg)} & \textbf{CO ($\frac{\text{L}}{\text{min}}$)}\\ \midrule\midrule
    1 & 1 & 64 & Male & 164.0 & 72.6 \ \ \ \ & 4.0\\
    2 & 4 & 58 & Male & 161.0 & 70.0 \ \ \ \ & 4.3\\
    3 & 1 & 27 & Female & 151.0 & 81.1 \ \ \ \ & 2.6\\
    4 & 4 & 71 & Female & 167.6 & 93.3 \ \ \ \ & 6.1\\
    5 & 4 & 51 & Male & 179.1 & 117.2 \ \ \ \ & 3.6\\
    6 & 1 & - & Male & 178.0 & 108.0 \ \ \ \ & 6.4\\
    7 & 1 & - & Male & 179.0 & 74.0 \ \ \ \ & 6.3\\
    8 & 1 & - & Female & 183.0 & 82.0 \ \ \ \ & 5.6\\
    9 & 4 & - & Female & 154.9 & 67.4 \ \ \ \ & 4.0\\ \midrule
    \end{tabular}\newline
\end{center}
{\footnotesize \hspace{2cm}CO: cardiac output; PH: pulmonary hypertension.}

{\footnotesize \hspace{2cm}For patients 6-9, age were omitted from medical records.}
\end{table}

\subsection{Data extraction}
\begin{figure}[ht!]
    \centering
    \includegraphics[width = 0.7\linewidth]{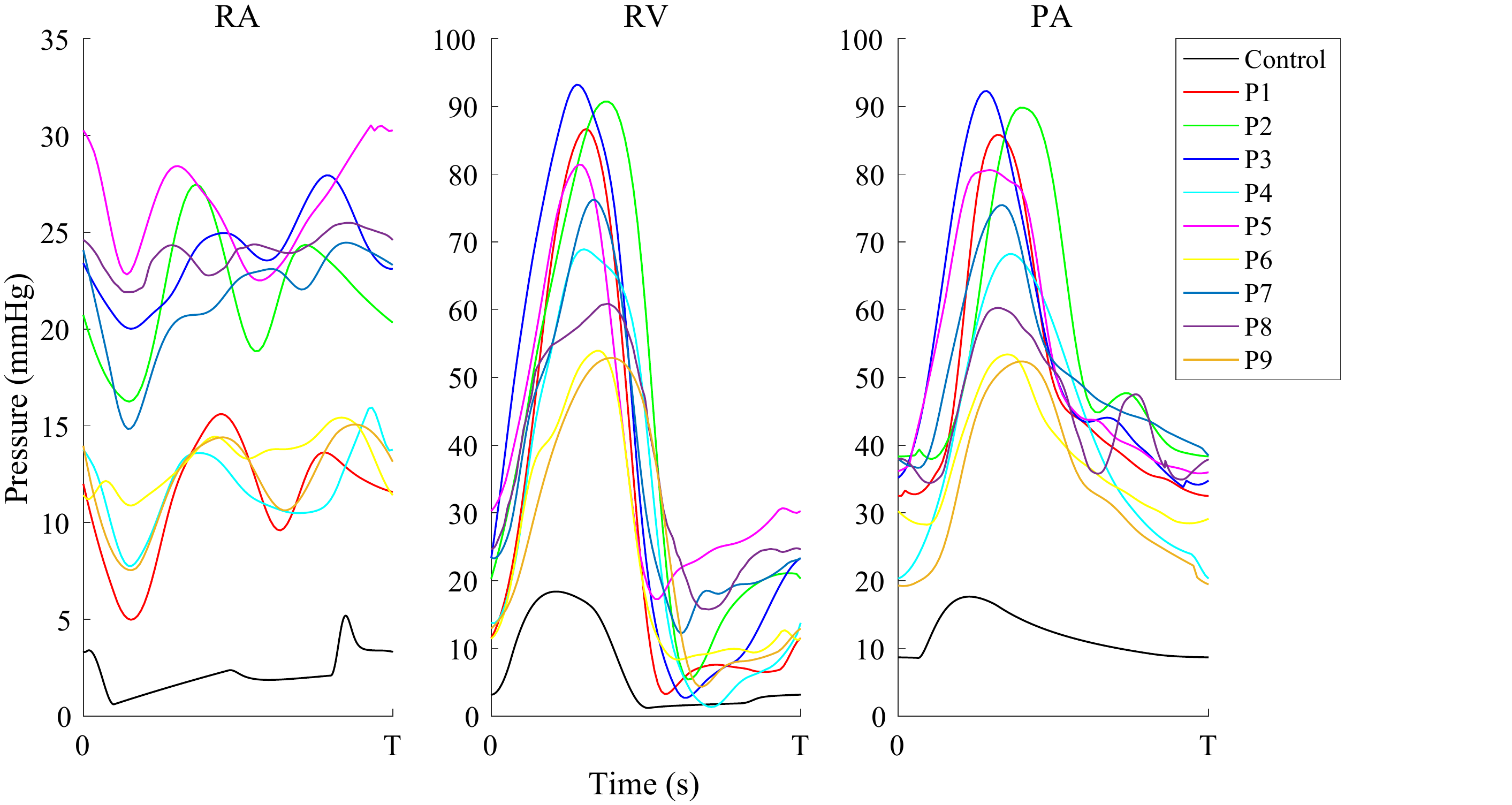}
    \caption{\textbf{Data processing.} Dynamic data from the right atrium (RA), right ventricle (RV), and main pulmonary artery (MPA) for each patient are digitized from right heart catheterization recordings and used for model calibration.}
    \label{fig:Data}
\end{figure}

Time-series data are extracted from clinical RHC reports using GraphClick version 3.0.3 for Mac OS and Map Digitizer available on the Apple AppStore. Beat-to-beat hemodynamic profiles for each patient are extracted by aligning RHC pressure waveform to the electrocardiogram signal. The waveforms are separated using by R-R interval and stored as separate files. For this study, a single representative RA, RV, and MPA signal is chosen for each patient (see figure \ref{fig:Data}). Since RHC data are not measured simultaneously, the representative waveforms are selected during expiration and assigned a cardiac cycle length equal to the averaged pressure cycle length. To align the signals within the cardiac cycle, we shift the RA and MPA signals to ensure that RA contraction occurs before the start of RV isovolumic contraction and that peak RV pressure occurs immediately before peak MPA pressure. Magnitudes of the RA, RV, and MPA pressure signals are shifted slightly to ensure physiological valve dynamics. Dynamic pressure waveforms from the RHC are shown in figure \ref{fig:Data}. Lastly, we construct a normotensive, control patient using pressure and volume values from literature \cite{Boron2017,KawelBoehm2015}; these pressure values are displayed in table \ref{tab:data}. Control parameters and model predictions are compared to those obtained using PH data. 

 \begin{table}[!htp] \centering
   \caption{Static values from patient data and used for nominal parameter calculations. Mean and standard deviation values are calculated for PH data only. † control values obtained from \cite{Boron2017,KawelBoehm2015}. ‡ left atrial diastolic value used in place of PAWP.} \label{tab:data}
    \centering
    \begin{tabular}{l|r|rrrrrrrrrc} \midrule
    \textbf{Data} & \textbf{Control} & \textbf{P1} & \textbf{P2} & \textbf{P3} & \textbf{P4} & \textbf{P5} & \textbf{P6} & \textbf{P7} & \textbf{P8} & \textbf{P9} & \textbf{Mean $\pm$ SD} \\ \midrule\midrule
    $p^d_{ra,M}$ & 12 \ \ \ \ & 14  & 24  & 28  & 16  & 31  & 15  & 24  & 25  & 15  & $21\pm6$\\
    $p^d_{ra,m}$ & 3  \ \ \ \  & 5   & 16  & 20  & 8   & 23  & 11  & 15 & 22  & 8   & $14\pm7$\\
    $p^d_{rv,M}$ & 21 \ \ \ \  & 87  & 91  & 93  & 69  & 81  & 54  & 76  & 61  & 53  & $74\pm15$\\
    $p^d_{rv,m}$ & 2 \ \ \ \   & 3   & 5   & 3   & 1   & 17  & 8   & 12  & 16  & 4   & $8\pm6$\\
    $p^d_{pa,M}$ & 21 \ \ \ \  & 86  & 90  & 92  & 68  & 81  & 53  & 75  & 60  & 52  & $73\pm15$\\
    $p^d_{pa,m}$ & 8  \ \ \ \  & 32  & 38  & 34  & 20  & 36  & 28  & 37  & 34  & 19  & $31\pm7$\\
    $p^d_{pa}$   & 12 \ \ \ \  & 48  & 55  & 54  & 41  & 53  & 37  & 51  & 45  & 34  & $46\pm8$\\
    $p^d_{W}$    & 5 ‡ \ \ \ \ & 4   & 5   & 8   & 11  & 20  & 10  & 17  & 22  & 12  & $12\pm6$\\
    $p^d_{sa,M}$ & 120 \ \ \ \  & 112 & 112 & 127 & 148 & 118 & 133 & 127 & 89  & 123 & $121\pm16$\\
    $p^d_{sa,m}$ & 80 \ \ \ \  & 76  & 76  & 90  & 78  & 77  & 87  & 92  & 68  & 65  & $79\pm9$\\
    $p^d_{sa}$   & 93 \ \ \ \  & 88  & 88  & 102 & 101 & 91  & 102 & 103 & 75  & 84  & $93\pm10$\\\midrule
    \end{tabular}
    \end{table} 

\subsection{Mathematical model}

\begin{figure}[ht!]
    \centering
    \includegraphics[width = 0.6\linewidth]{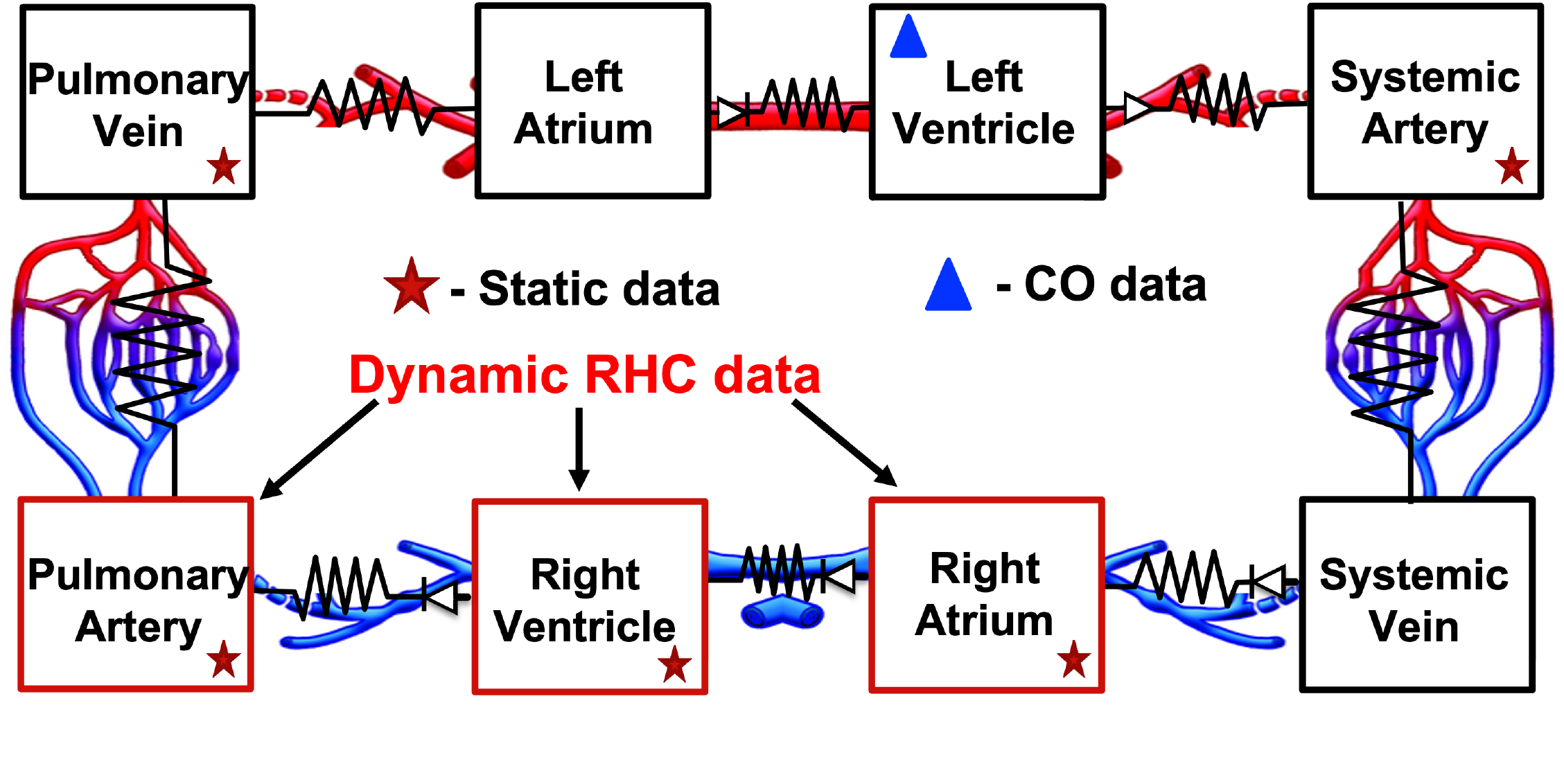}
    \caption{\textbf{Model schematic.} Follows an electrical circuit analog. The model has eight compartments: the systemic and pulmonary arteries and veins, two atria, and two ventricles. Each compartment is modeled as compliant and is separated by a resistor element. The right atrium, right ventricle, and pulmonary arteries (red boxes) have both dynamic and static data. The pulmonary veins and systemic arteries have only static data. RHC: right heart catheterization; CO: cardiac output.}
    \label{fig:Model}
\end{figure}

This study utilizes a systems-level, ordinary differential equations (ODE) model (shown in figure \ref{fig:Model}) that simulates dynamic pressure $p$ (mmHg), flow $q$ (mL/s), and volume $V$ (mL). The model consists of 8 compartments: the left and right atria and ventricles, and the systemic and pulmonary arteries and veins. The model is formulated using an electrical circuit analogy, with pressure analogous to voltage, flow to current, volume to charge, and compliance to capacitance. We include four heart valves, two semilunar (tricuspid and mitral), and two atrioventricular (pulmonary and aortic). An additional systemic venous valve is also included. To ensure proper flow between compartments, heart valves are modeled as diodes, i.e., the valves are either open or closed depending on the pressure gradient between compartments. Equivalent to an RC circuit, equations relating to the three dependent quantities are given by 
\begin{eqnarray}
    \frac{dV_{s,i}}{dt} &=& q_{i-1} - q_i, \label{eqn:1} \\
   q_i &= &\frac{p_i - p_{i+1}}{R_i}, \label{eqn:2}\\
    V_{s,i} &=& V_i - V_{un,i} = C_i(p_i -p_{i+1}), \label{eqn:3}
\end{eqnarray}
where the subscripts $i-1$, $i$, $i+1$ refer to the prior, current, and proceeding compartments in the system, respectively. $V_{s,i}$ (mL) denotes the stressed volume (the circulating volume), and $V_{un,i}$ (mL) is the unstressed volume (the non-circulating volume, assumed constant). $R_i$ (mmHg$\cdot$s/mL) denotes the resistance between two compartments and $C_i$ (ml/mmHg) the compartment compliance. Equation \eqref{eqn:1} ensures conservation of volume, equation \eqref{eqn:2} is the analog of Ohm’s law, and equation \eqref{eqn:3} relates volume and pressure.

\begin{figure}[ht!]
    \centering
    \includegraphics[width = 0.35\linewidth]{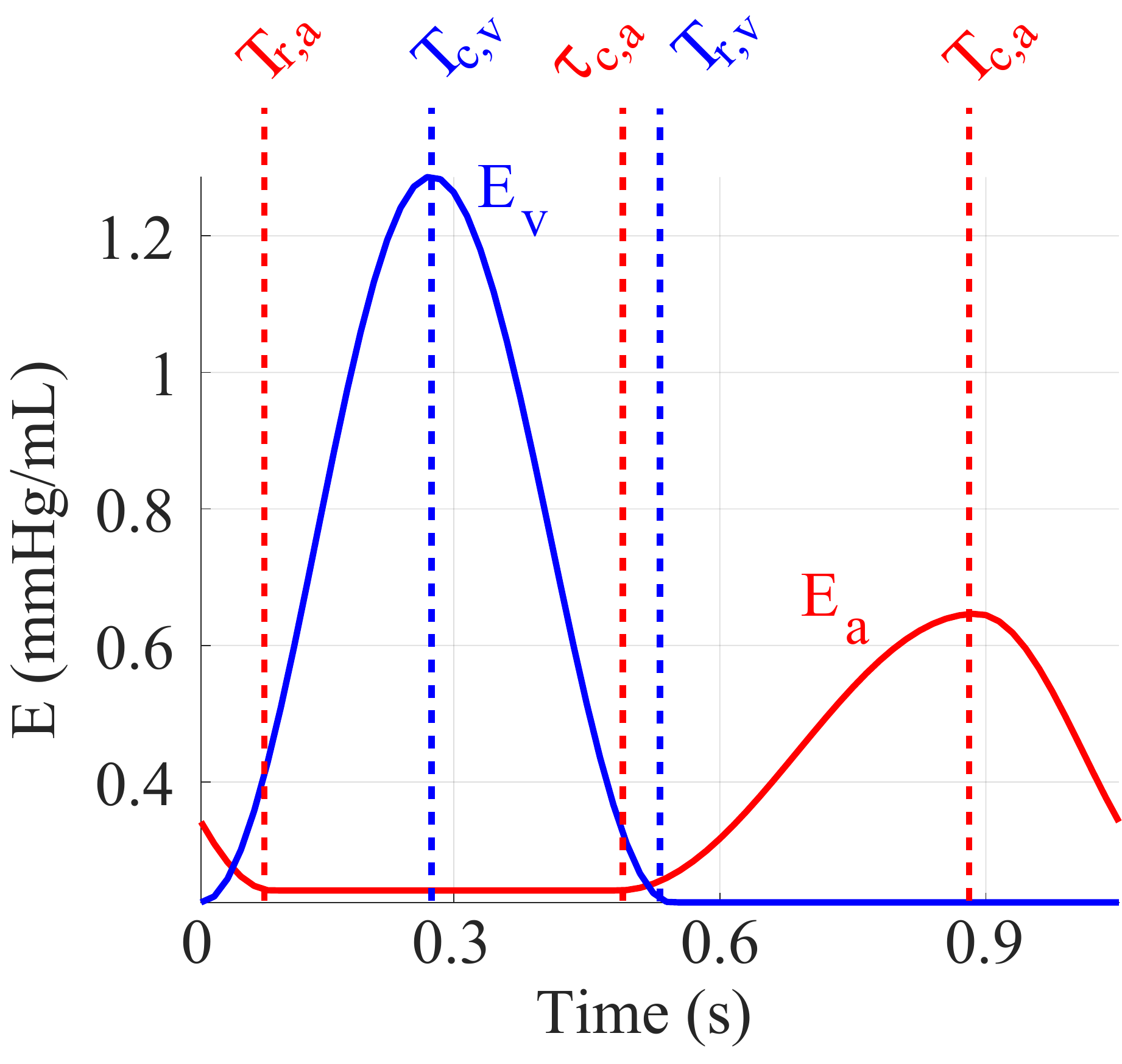}
    \caption{\textbf{Heart chamber elastance function.} Representative elastance function for the atrial (red) and ventricular (blue) heart chambers. Timing parameters are shown above their respective phases of the cardiac cycle. Note that ventricular isovolumic contraction occurs while the atrium is still relaxing.}
    \label{fig:Elast}
\end{figure}

We model each heart chamber by a time-varying elastance function $E_i(t)$  (mmHg/mL) \cite{Ellwein2008,Marquis2018}, which relates pressure and volume by
\begin{equation}
    p_i\left(t\right)=E_i\left(\tilde{t}\right)V_{s,i},
    \label{eqn:4}
\end{equation}
where $i=ra,la,rv,lv$ denote the left $(l)$ and right $(r)$ atria $(a)$ and ventricles $(v)$. The time within the cardiac cycle is denoted by $\tilde{t}=\mathrm{mod}(t,T$), where $T(s)$ is the length of the cardiac cycle. The ventricular elastance function $E_v(\tilde{t})$ is given by the piece-wise continuous function \cite{Ellwein2008}
\begin{align}
E_v(\tilde{t}) = 
\begin{cases}
\frac{E_{M}-E_{m}}{2} \Big(\cos\Big(\frac{\pi\tilde{t}}{T_{c}}\Big)\Big) +E_{m} , &  0\le \tilde{t} \le T_{c} \\ 
\frac{E_{M}-E_{m}}{2}\Big(1+\cos\Big(\frac{\pi\big(\tilde{t}-T_{c}\big)}{\big(T_{r} - T_{c}\big)}\Big)\Big) +E_{m} , & T_{c} < \tilde{t} \le T_{r} \\ 
E_{m}, & T_{r} < \tilde{t} \le T,
\end{cases} 
\label{eqn:5}
\end{align}
where $E_{v,m}$ and $E_{v,M}$ (mmHg/mL) are the minimal and maximal ventricular elastances, and $T_{c,v}$ (s) and $T_{r,v}$ (s) denote the duration of ventricular contraction and relaxation. The atrial elastance function (shown in figure \ref{fig:Elast}) is prescribed in a similar fashion \cite{Liang2009}
\begin{align}
E_a(\tilde{t})=  
\begin{cases}
\frac{E_{a,M}-E_{a,m}}{2} \Big(1-\cos\Big(\frac{\pi\big(\tilde{t}- T_{r,a}\big)}{\big(T - T_{c,a} + T_{r,a}}\Big)\Big) +E_{a,m} , &  0\le \tilde{t} \le T_{r,a} \\ 
E_{a,m}, & T_{r,a} < \tilde{t} \le \tau_{c,a} \\ 
\frac{E_{a,M}-E_{a,m}}{2}\Big(1-\cos\Big(\frac{\pi\big(\tilde{t}-\tau_{c,a}\big)}{\big(T_{c,a} - \tau_{c,a}\big)}\Big)\Big) +E_{a,m} , & \tau_{c,a} < \tilde{t} \le T_{c,a} \\ 
\frac{E_{a,M}-E_{a,m}}{2}\Big(1+\cos\Big(\frac{\pi\big(\tilde{t}-T_{c,a}\big)}{\big(T - T_{c,a} + T_{r,a}\big)}\Big)\Big) +E_{a,m} , & T_{c,a} < \tilde{t} \le T. 
\end{cases} 
\label{eqn:6}
\end{align}
Here, $E_{a,m}$ and $E_{a,M}$ (mmHg/mL) are the minimum and maximum elastances of the atria and $T_{r,a}$, $\tau_{c,a}$ and $T_{c,a}$ (s) denote the start of atrial relaxation, the start of atrial contraction, and the point of maximum atrial contraction. The elastance model is parameterized by $0\le T_{r,a}\le\ \tau_{c,a}\le T_{c,a}\le T$. Figure \ref{fig:Elast} shows a representative elastance time course in the atria and ventricles.

\subsection{Model outcomes}
We compute four physiological quantities derived from the model predictions and inferred parameters. These indices are utilized as biomarkers of PH severity. 
\begin{enumerate}
    \item \textbf{Stroke work per cycle (SW):} defined as the time-averaged integral of the PV loop, i.e., $\displaystyle \frac{1}{T}\int_{V}^{\ }p(t)dV'$, calculated in each heart chamber \cite{Rimehaug2013,Colunga2020}.
    \item \textbf{Resistance ratio:} the ratio of pulmonary and systemic resistance, $R_p/R_s$ \cite{Yang2018}.
    \item \textbf{Compliance ratio:} the ratio of pulmonary and systemic compliance, $C_{pa}/C_{sa}$.
    \item \textbf{Pulsatility index (PI):} the ratio of pulmonary arterial pulse pressure to average right atrial pressure, $(p_{pa,M}-p_{pa,m})/{\bar{p}}_{ra}$ \cite{Mazimba2019}.
\end{enumerate}
    
\subsection{Parameter values and initial conditions}
The sparse hemodynamic data and many model parameters make it imperative that nominal parameter values and initial conditions are set in a physiologically and patient-specific manner. Following previous approaches \cite{Marquis2018,Colunga2020}, we use a combination of patient-specific data (where available) and literature values. Table \ref{tab:A1} lists the nominal parameter values and their calculation.  

\subsubsection{Compartment volumes and cardiac output} \label{sec:volumes}
Using Hidalgo’s formula \cite{Hidalgo1962}, each patients' total blood volume ($V_{tot}$, mL) is calculated as a function of height ($H$, cm), weight ($W$, kg), and sex \cite{Williams2019} as
\begin{equation}
V_{tot}= 
    \begin{cases}
    3.47\cdot \text{BSA}-1.954\cdot1000, & \text{if Female} \\ 
    3.29\cdot \text{BSA}-1.229\cdot1000, & \text{if Male}   
    \end{cases},
    \label{eqn:7}
\end{equation}
where BSA$=\sqrt{W\cdot H/3600}$ (m$^2$) is the body surface area \cite{DeCherney1982}.

The heart's initial stressed volumes (initial conditions) are calculated using BSA indexed values. In contrast, stressed volumes in the vasculature are based on blood volume proportions \cite{Beneken1967}. The BSA indexed volumes, $V_{i,ED}^d$, for the right heart are based on Tello et al. \cite{Tello2019}, with $V_{ra,ED}^d=58.9 \cdot BSA$ and  $V_{rv,ED}^d=116.9 \cdot BSA$. We assume that the left heart chamber volume is unaffected by PH, and use $V_{la,ED}^d=30 \cdot BSA$ and  $V_{lv,ED}^d=80 \cdot BSA$ \cite{KawelBoehm2015}. Note that these values determine the blood volume distributions for PH patients. The normotensive control simulation used $V_{ra,ED}^d=30 \cdot BSA$ and  $V_{rv,ED}^d=78 \cdot BSA$, $V_{la,ED}^d=30 \cdot BSA$, and  $V_{lv,ED}^d=78 \cdot BSA$ \cite{KawelBoehm2015}.

The total volumes for the systemic and pulmonary arteries are 13\% and 3\% of $V_{tot}$, of which the stressed volumes are 27\% and 58\% of the total volume. Pulmonary venous blood volume is 11\% of $V_{tot}$, and 11\% of this volume is stressed. These values are from previous studies \cite{Ellwein2008,Marquis2018}. To ensure that blood volume distributions add to 100\%, we calculate total systemic venous blood volume as the remaining volume
\begin{equation*}
    V_{sv\%} = 100 - 13 - 3 - 11 - V_{H\%},
\end{equation*}
where $V_{H\%}$ is the percentage of total blood volume within the entire heart. CO is calculated assuming that the total blood volume circulates in one minute \cite{Ellwein2008,Boron2017}. 

\subsubsection{Pressure} Pulmonary circulation pressures are extracted from the RHC data, while systemic arterial pressure is determined from cuff measurements. These values are listed in table \ref{tab:data}. Nominal pressure values for compartments we do not have measurements (i.e., the left atrium, LV, and systemic veins) are calculated by scaling pressures in their adjacent, data calibrated compartments \cite{Williams2019}. We use the following relationships for compartments for which we do not have data
\begin{eqnarray}
    p_{sv}&=&\max(10,1.15\ p_{rv,m}),
    \label{eqn:8}\\
    p_{la,m}&=&0.95\ p_{pv},
    \label{eqn:9}\\
    p_{la,M}&=&p_{la,m}+5,
    \label{eqn:10}\\
    p_{lv,m}&=&0.97{\ p}_{la,M},
    \label{eqn:11}\\
    p_{lv,M}&=&1.01\ p_{sa,M}.
    \label{eqn:12}
\end{eqnarray}
The subscripts $sa,sv,la,$ and $pv$ denote the systemic arteries, systemic veins, left atrium, and pulmonary veins, respectively. The additional subscript $m$ and $M$ denote minimum and maximum value. For the left atrium, we assume a pulse pressure of 5 mmHg, consistent with previous studies \cite{Pironet2013}.

\subsubsection{Resistance} Each compartment is separated by a resistance to flow. Utilizing Ohm’s law, the nominal vascular resistance is calculated as
\begin{equation}
    R_i=\frac{\Delta p}{\text{CO}},	
    \label{eqn:13}
\end{equation}
where the resistance in compartment $i$ depends on the pressure gradient, $\Delta p$, and the CO; refer to table \ref{tab:A1} for more details. The the aortic and pulmonary valve resistances are calculated as
\begin{equation}
    R_{ava}=\frac{p_{lv,M}\ -p_{sa,M}}{\text{CO}}\ \ \ \text{and}\ \ R_{pva}=\frac{p_{rv,M}\ -p_{pa,M}}{\text{CO}},
    \label{eqn:14}
\end{equation}
For PH patients, RA and pulmonary venous pressures are elevated \cite{Alenezi2020} and resistance equations overestimate atrioventricular valve resistance. To circumvent this, we set $R_{tva}=0.03$ and $R_{mva}=0.01$ for all nine PH patients. 

\subsubsection{Compliance} is defined as the relative change in volume for a given change in pressure \cite{Wang2011} and quantifies the ability of the vasculature to distend under load. In this study, nominal compliance estimates are
\begin{equation}
    C_i=\frac{V_i - V_{un,i}}{\tilde{p}_{i}},
    \label{eqn:16}
\end{equation}
where $\tilde{p}_{i}$ is a compartment specific pressure \cite{Colunga2020}; see table \ref{tab:A1} for more details.

\subsubsection{Heart parameters} include elastance and timing parameters. Noting that compliance is the inverse of elastance and that the compliance in the heart is minimal during end-systole (computed at the maximum pressure and minimal volume) \cite{Marquis2018}, we calculate the maximum and minimum elastances as
\begin{equation}
    E_{i,M}=\frac{p_{i,M}}{V_{i,m} - V_{un,i}}\ \ \ \mbox{and}\ \ \ \ E_{i,\ m}=\frac{p_{i,m}}{V_{i,M} - V_{un,i}},	
    \label{eqn:17}
\end{equation}
where $i=la, ra, lv, rv$.

Nominal timing parameters for the RA and RV elastance functions are extracted from the time-series data. Maximum and minimum RV elastance occur at peak systole and the beginning of diastole, corresponding to $T_{c,v}$ and $T_{v,r}$, respectively. RA dynamics are used to determine the end of atrial systole, the start of atrial contraction, and peak atrial contraction, i.e. $T_{r,a}, \tau_{c,a},\text{ and }\ T_{c,a}$. Since dynamic data is unavailable for the left atrium and LV, we set left-heart chamber timing parameters equal to the right-heart timing parameters. 

\begin{table}[!htp]
\begin{center}
\caption{Parameters in the 0D model and the methods for calculating their nominal values.}\label{tab:A1}
\footnotesize
\begin{tabular}{cccc}\midrule
{\bf Parameter} & {\bf Units} & {\bf Equation} & {\bf Reference} \\ \midrule\midrule
\multicolumn{4}{c}{\textbf{Heart Valves}} \\ \midrule
$R_{ava}$ & $\frac{mmHg\ s}{mL}$ & 
$\frac{\displaystyle  p_{lv,M}-p_{sa,M}}{\displaystyle q_{tot}}$ &  Ohm's Law \\ 
$R_{mva}$ & $\frac{mmHg \ s}{mL}$ & 0.01 & - \\
$R_{pva}$ & $\frac{mmHg \ s}{mL}$ & 
$\frac{\displaystyle p_{rv,M}-p_{pa,M}}{\displaystyle q_{tot}}$    & Ohm's Law \\
$R_{tva}$ & $\frac{mmHg \ s}{mL}$ &  0.03  & - \\
$R_{sv}$  & $\frac{mmHg \ s}{mL}$ & 
$\frac{\displaystyle \bar{p}_{sv}-p_{ra,m}}{\displaystyle q_{tot}}$ & Ohm's Law \\
$R_{pv}$  & $\frac{mmHg \ s}{mL}$ & 
$\frac{\displaystyle \bar{p}_{pv}-p_{la,m}}{\displaystyle q_{tot}}$ & Ohm's Law \\ \midrule\midrule
\multicolumn{4}{c}{\textbf{Systemic Vasculature}} \\ \midrule
$R_{s}$   & $\frac{mmHg \ s}{mL}$ & 
$\frac{\displaystyle p_{sa,m}-\bar{p}_{sv}}{\displaystyle q_{tot}}$ & Ohm's Law \\ 
$C_{sa}$  & $\frac{mL}{mmHg}$      & 
$\frac{\displaystyle V_{sa,M} - V_{sa,un}}{\displaystyle p_{sa,m}}$  &\cite{Colunga2020}\\
$C_{sv}$ & $\frac{mL}{mmHg}$ & 
$\frac{\displaystyle V_{sv,M}- V_{sv,un}}{\displaystyle \bar{p}_{sv}}$&\cite{Colunga2020}\\  \midrule\midrule
\multicolumn{4}{c}{\textbf{Pulmonary Vasculature}} \\ \midrule
$R_{p}$   & $\frac{mmHg \ s}{mL}$ 
& $\frac{\displaystyle p_{pa,m}-\bar{p}_{pv}}{\displaystyle q_{tot}}$ &  Ohm's Law\\
$C_{pa}$  & $\frac{mL}{mmHg}$ & 
$\frac{\displaystyle V_{pa,M} - V_{pa,un}}{\displaystyle p_{pa,m}}$ &\cite{Colunga2020}\\ 
$C_{pv}$ & $\frac{mL}{mmHg}$ & 
$\frac{\displaystyle V_{pv,M}- V_{pv,un}}{\displaystyle \bar{p}_{pv}}$&\cite{Colunga2020}\\ \midrule\midrule
\multicolumn{4}{c}{\textbf{Heart Elastance}}\\ \midrule
$E_{M,rv}$ & $\frac{mmHg}{mL}$ & 
$\frac{\displaystyle p_{rv,M}}{\displaystyle V_{rv,m} - V_{rv,un}}$&\cite{Marquis2018}\\
$E_{m,rv}$ & $\frac{mmHg}{mL}$ & 
$\frac{\displaystyle p_{rv,m}}{\displaystyle V_{rv,M} - V_{rv,un}}$ &\cite{Marquis2018}\\
$E_{M,ra}$ & $\frac{mmHg}{mL}$ & 
$\frac{\displaystyle p_{ra,M}}{\displaystyle V_{ra,m}-V_{ra,un}}$ &\cite{Marquis2018}\\
$E_{m,ra}$ & $\frac{mmHg}{mL}$ & 
$\frac{\displaystyle p_{ra,m}}{\displaystyle V_{ra,M}-V_{ra,un}}$ &\cite{Marquis2018}\\
$E_{M,lv}$ & $\frac{mmHg}{mL}$ & 
$\frac{\displaystyle p_{lv,M}}{\displaystyle V_{lv,m}-V_{lv,un}}$&\cite{Marquis2018}\\
$E_{m,lv}$ & $\frac{mmHg}{mL}$ & 
$\frac{\displaystyle p_{lv,m}}{\displaystyle V_{lv,M}-V_{lv,un}}$&\cite{Marquis2018}\\
$E_{M,la}$ & $\frac{mmHg}{mL}$ & 
$\frac{\displaystyle p_{la,M}}{\displaystyle V_{la,m}-V_{la,un}}$&\cite{Marquis2018}\\
$E_{m,la}$ & $\frac{mmHg}{mL}$ & 
$\frac{\displaystyle p_{la,m}}{\displaystyle V_{la,M}-V_{la,un}}$&\cite{Marquis2018}\\ \midrule\midrule
\multicolumn{4}{c}{\textbf{Heart Timing}}\\ \midrule
$\tau_{r,a}$ & $s$ & Data & - \\
$T_{c,a}$ & $s$ & Data & - \\
$T_{r,a}$ & $s$ & Data &  - \\
$T_{c,v}$ & $s$ & Data & - \\
$T_{r,v}$ & $s$ & Data & - 
\end{tabular}

\end{center}
\end{table}

\subsection{Model summary}

The model consists of a system of eight ODE’s, a stressed volumes, $V_{s,i}$, for each compartment, with twenty-five parameters. The system can be written as
\begin{align}
\mathbf{y}&=g(t,\mathbf{x};\theta),\\ \nonumber
\frac{d\mathbf{x}}{dt}&=f(t,\mathbf{x};\theta),\\ \nonumber
\mathbf{x}&=\{V_{la},V_{lv},V_{sa},V_{sv},V_{ra},V_{rv},V_{pa},V_{pv}\},
\label{eqn:18}
\end{align}
where
\begin{equation}
\begin{split}
\theta=\{R_s,R_p,R_{ava},R_{mva},R_{pva},R_{tva},R_{pv},R_{sv},C_{sa},C_{sv},C_{pa},C_{pv}, \\
E_{la,M},E_{la,m}, E_{ra,M},E_{ra,m},E_{lv,M},E_{lv,m}, E_{rv,M},E_{rv,m},\\
T_{r,a},\tau_{c,a},T_{c,a}, T_{c,v},T_{r,v},\}.
\end{split}
\label{eqn:18b}
\end{equation}
Here $\mathbf{x}$ denotes the state variables ($V_{s,i}$ in compartment $i$). The functions $f(t,\mathbf{x};\theta)$ denote the evolution of the states (equation \eqref{eqn:1}), and $\bm{\theta}$ are the parameters. The vector $\mathbf{y}$ is the model output, including predictions of pressure and CO, used for parameter inference. 

\subsection{Parameter inference}

We estimate model parameters, some of which correspond to disease biomarkers, by minimizing the relative least-squares error between model predictions and data. We use the Levenberg-Marquardt algorithm to solve the generalized least-squares problem \cite{Kelley1999}. The observed data $\mathbf{y}^d$ (static or time-series) is assumed to be of the form
\begin{equation}
    \mathbf{y}^d=g(t,\mathbf{x};\mathbf{\theta})+\mathbf{\varepsilon},
    \label{eqn:19}
\end{equation}
where $g(t,\mathbf{x};\theta)$ are the model predictions (here, pressure and CO), and $\varepsilon$ is the measurement error, assumed to be independent and identically distributed (iid) white Gaussian noise, i.e., $\varepsilon\ \sim \ \mathcal{N}(0,\ \sigma_\varepsilon^2\mathbf{I})$. Using this framework, we estimate parameters that minimize the relative sum of squared errors, $J=\mathbf{r}^T\mathbf{r}$, where $\mathbf{r}$ is the residual vector. The residual encompasses the relative differences between the measured data $\mathbf{y}^d$ and model predictions\ $\mathbf{y}=\ g(t,\mathbf{x};\mathbf{\theta})$. 

The static residual is defined as
\begin{equation}
    \mathbf{r}_s=\frac{1}{\sqrt{N_s}}\frac{\mathbf{y}-\mathbf{y}^d}{\mathbf{y}^d},	
    \label{eqn:20}
\end{equation}
where the vector $\mathbf{y}=[p_{ra,M},\ p_{ra,m},\ p_{rv,M},\ p_{rv,m},\ p_{pa,M},\ p_{pa,m},$ $\ p_{sa,M},\ p_{sa,m},\ p_{pv,m},\ \text{CO}]$ includes model outputs, $\mathbf{y}^d$ is the corresponding data, and $N_s$ is the number of points. The three dynamic residuals are given by
\begin{eqnarray}
    \mathbf{r}_{ra}&=&\frac{1}{\sqrt{N_{ra}}}\frac{\mathbf{p}_{ra}(t;\theta)-\mathbf{p}_{ra}^d(t)}{\mathbf{p}_{ra}^d(t)},
    \label{eqn:21}\\
    \mathbf{r}_{rv}&=&\frac{1}{\sqrt{N_{rv}}}\frac{\mathbf{p}_{rv}(t;\theta)-\mathbf{p}_{rv}^d(t)}{\mathbf{p}_{rv}^d(t)},
    \label{eqn:22}\\
    \mathbf{r}_{pa}&=&\frac{1}{\sqrt{N_{pa}}}\frac{\mathbf{p}_{pa}(t;\theta)-\mathbf{p}_{pa}^d(t)}{\mathbf{p}_{pa}^d(t)},	
    \label{eqn:23}
\end{eqnarray}
where $\mathbf{p}_i(t;\theta),\ \mathbf{p}_i^d(t),$ and $N_i$ are the time-series pressure predictions, time-series pressure data, and number of residual points for the RA, RV, and pulmonary arteries. we consider two combined residuals as our quantity of interest
\begin{flalign*}
\begin{split}
    \mathbf{r}_\mathbf{1}\ &=\ \mathbf{r}_s, \\
\mathbf{r}_\mathbf{2}&=[\mathbf{r}_s,\ \mathbf{r}_{ra},\ \mathbf{r}_{rv},\ \mathbf{r}_{pa}].
\end{split}
\end{flalign*}
Similar to the approach in \cite{Marquis2018}, each residual is computed over the last 30 cycles of the model predictions compared to the data. 
In the absence of volume data, we include four penalty terms in our inference procedure to constrain heart chamber volumes. PAH and CTEPH patients have enlarged RAs and RVs, increasing the chamber volume \cite{Tello2019}. We penalize end-diastolic model predictions below a BSA-indexed volume threshold, as defined in subsection \ref{sec:volumes}. The penalty functions are defined by
\begin{equation}
    J_{\text{penalty},i} = \max\left(0,\frac{\max(\mathbf{V}_i)-V_{i,ED}^d}{V_{i,ED}^d}\right),
\end{equation}
where $i=la,lv,ra,rv$ and$\mathbf{V}_i$ is the predicted chamber volume.

\subsection{Sensitivity analyses}
We compute the sensitivity of the residual vectors $\mathbf{r}_\mathbf{1}$ and $\mathbf{r}_\mathbf{2}$ with respect to the model parameters. Both local, derivative-based, and global, variance-based, sensitivity analyses are used. The former methods are valid within a small neighborhood of the nominal parameter values and quantify the gradient of the residual vectors $\mathbf{r}_\mathbf{1}$ and $\mathbf{r}_\mathbf{2}$ with respect to the parameters. The latter measure model sensitivity throughout the physiological parameter space, simultaneously varying multiple factors. 
 
The local sensitivity of the residual for a parameter $\theta_i$ at time $t$ is denoted by $\chi_{i}(t)$. Sensitivities are approximated numerically via the complex-step method \cite{DePauw2005}. We rank parameters from most to least influential by calculating the 2-norm of each sensitivity \cite{Marquis2018,Colunga2020}
\begin{equation}
    \| \chi_i(t)\|_2^2 = \Bigg(\sum_{l=1}^N\chi_i^2(t_l)\Bigg)^{\frac{1}{2}},
    \label{eqn:R}
\end{equation}
where $i = 1,2,\dots, \mathcal{M}$ is the number of parameters and $l = 1,2,\dots, N$ is the length of the time vector.

While global sensitivity analysis is more computationally expensive than local methods, its ability to vary multiple parameters at a time may expose undiscovered relationships between parameters \cite{Eck2016}. In this study, we use variance-based global sensitivity analysis methods, computing first ($S$) and total order ($S_T$) Sobol’ indices \cite{Sobol2001}). The former measures the parameters’ individual contribution to the total output variance of the cost function, and the latter the individual contributions and higher-order interactions between the parameters on the variance.  $S_T$ are used to order parameters from most to least influential. Additional local and global methods information can be found in Section S2 of the Supplemental Material.

\subsection{Parameter subset selection}
Once the sensitivity analysis is performed, additional steps are taken to determine if the parameters are identifiable. A parameter set to be identifiable if it can be uniquely determined when fitting a model to data \cite{Pope2009,Mader2015}. The model used in this study is analogous to an electrical resistor-capacitor circuit. Circuit theory dictates that resistors and capacitors in series and parallel can be combined to give an equivalent resistor and capacitor. Therefore, if no data is available between two components, their parameters cannot be estimated uniquely, i.e., they are non-identifiable. Given the limited data and the large number of parameters (found in \eqref{eqn:18b}), we expect identifiability problems if all parameters are inferred from data \cite{Marquis2018,Colunga2020}. We take several steps to determine an identifiable and influential subset with respect to the residual vectors. The subset selection process begins by analyzing the global sensitivity results. Parameters with $S_{T_i}\approx0$ are considered non-influential and fixed at their nominal values \cite{Sumner2012,Eck2016}. After excluding these parameters, we use a singular value decomposition (SVD) QR factorization method to determine local pairwise parameter interactions \cite{Pope2009}. Lastly, we use multistart inference to reduce the subset further until we mitigate all identifiability issues.

\subsubsection{SVD-QR} The SVD-QR method \cite{Golub1976} decomposes the sensitivity matrix as $\bm{\chi} = \bm{U}\bm{\Sigma} \bm{V}^\top$, where $\bm{U}$ is the matrix of left orthonormal eigenvectors, $\bm{\Sigma}$ is a diagonal matrix of singular values, and $\bm{V}$ is the matrix of right orthonormal eigenvectors. The total number of identifiable parameters, $\rho$, is the numerical rank of $\Sigma$ and is used to partition $\bm{V}$ as $\bm{V} = [\bm{V}_\rho \ \ \bm{V}_{P-\rho}]$. The permutation matrix $\tilde{\bm{P}}$ is determined by QR factorization such that $\bm{V}_\rho^\top\tilde{\bm{P}} = \bm{Q}\bm{R}$. Here, $\bm{Q}$ is an orthogonal matrix, and the first $\rho$ columns of $\bm{R}$ form an upper triangular matrix consisting of diagonal entries in decreasing order. The first $\rho$ entries of $\tilde{\bm{P}}$ establish the identifiable parameters for the subset.

\subsubsection{Multistart inference} The previous methods ensure that the parameters are locally and linearly identifiable. However, they do not guarantee practically identifiable parameter subsets if the model has nonlinear behavior in output space \cite{Sumner2012}. Thus, we determine our final subset by inferring parameters from multiple initial guesses randomly selected between $\pm 20\%$ of the nominal values. Non-identifiable parameters likely approach different values, whereas identifiable parameters converge to the same value regardless of initial guess \cite{Schiavazzi2017}. We assess identifiability by calculating each patient's coefficient of variance (CoV; the standard deviation relative to the mean). Subsets that exhibit parameter CoV $>10\%$ are reduced by fixing the least influential parameter above this threshold. The multistart inference is iteratively run until the CoV for each parameter is below the $10\%$ threshold.

\subsection{Confidence and prediction intervals}
Model parameter and output uncertainty are quantified using asymptotic analysis \cite{Colebank2019}. Under the assumption that the  noise $\mathbf{\varepsilon}$ is iid, we compute the variance estimator $\hat{\sigma}_\epsilon^2$ and parameter covariance estimator $\hat{\mathbf{C}}=\hat{\sigma}_\epsilon^2\left(\hat{\bm{\chi}}^\top\hat{\bm{\chi}}\right)^{-1}$ using asymptotic analysis for nonlinear least-squares \cite{Smith2013}.

The 95\% parameter confidence intervals for each inferred parameter, $\hat{\theta}_i$, are computed as
\begin{equation}
    [{{\hat{\theta}}_i^{CI-},\hat{\theta}}_i^{CI+}]={\hat{\theta}}_i\pm t_{N-\rho}^{0.975}\sqrt{{\hat{\mathbf{C}}}_{i,i}},
    \label{eqn:25}
\end{equation}
where $t_{N-\mathcal{M}^\prime}^{1-\alpha/2}$ is a two-sided t-statistic with a $1-\alpha/2=95\%$ confidence level, and $\sqrt{{\hat{\mathbf{C}}}_{i,i}}$ represents the standard error for the $i$th parameter estimator. Throughout we denote these confidence intervals by mean $\pm$ two standard deviations, i.e. $\hat{\theta_i}\pm 2\sigma_{\theta_i}$. The confidence and prediction intervals for the optimal model output ${\hat{y}}_j$ at time $t_j$ are
\begin{equation}
    [{{\hat{y}}_j^{CI-},\hat{y}}_j^{CI+}]={\hat{y}}_j\pm t_{N-\rho}^{0.975}\ \sqrt{\hat{\bm{\chi}}_{j}^T{\hat{\mathbf{C}}}_{i,i}\hat{\bm{\chi}}_{j}},	
    \label{eqn:26}
\end{equation}

\begin{equation}
    [{{\hat{y}}_j^{PI-},\hat{y}}_j^{PI+}]={\hat{y}}_j\pm t_{N-\rho}^{0.975}\ \sqrt{\sigma_\varepsilon^2+\hat{\bm{\chi}}_{j}^T{\hat{\mathbf{C}}}_{i,i}\hat{\bm{\chi}}_{j} },	
    \label{eqn:27}
\end{equation}
where $\hat{\bm{\chi}}_j^T$ is the sensitivity vector at $t_j$ evaluated at  $\hat{\mathbf{\theta}}=\left\{{\hat{\mathbf{\theta}}}_\mathbf{\rho},\mathbf{\theta}_{\mathcal{M}-\mathbf{\rho}}\right\}$. Note that the prediction intervals account for the variance in both the model output and the data, hence they are wider. 

\subsection{Simulations} 
To study the impact of PH, we run several simulations comparing PH patients to a normotensive control subject.

\paragraph{Control:} Simulations for a control patient are conducted using normotensive pressure and volume values given in table \ref{tab:data}. Hemodynamic predictions are compared to those from PH patients.

\paragraph{Static:} Similar to Colunga et al. \cite{Colunga2020}, we calibrate model predictions utilizing only static pressure and CO data for each PH patient, i.e., $\mathbf{r}_1$. We use this as a benchmark procedure to determine the effects of adding dynamic waveforms.

\paragraph{Dynamic waveforms:} Model predictions of systolic, diastolic, and mean pressure are computed in combination with dynamic RA, RV, and pulmonary artery predictions utilizing residual $\mathbf{r}_2$. 

\section{Results}
Local and global sensitivity analyses of both residuals $\mathbf{r}_1$ and $\mathbf{r}_2$ distinguish influential and non-influential parameters. Next, SVD-QR and multistart inference are used to construct subsets of identifiable parameters. Model predictions are calibrated to measured RHC data using the identifiable subset, and other outcomes, such as PV loops, are computed. Uncertainty of parameter estimates and model outputs are compared for the two residual vectors and is shown here for a single representative patient; results for remaining patients can be found in the Supplemental Material.

\subsection{Sensitivity analyses}
\begin{figure}[ht!]
    \centering
    \includegraphics[width=0.8\columnwidth]{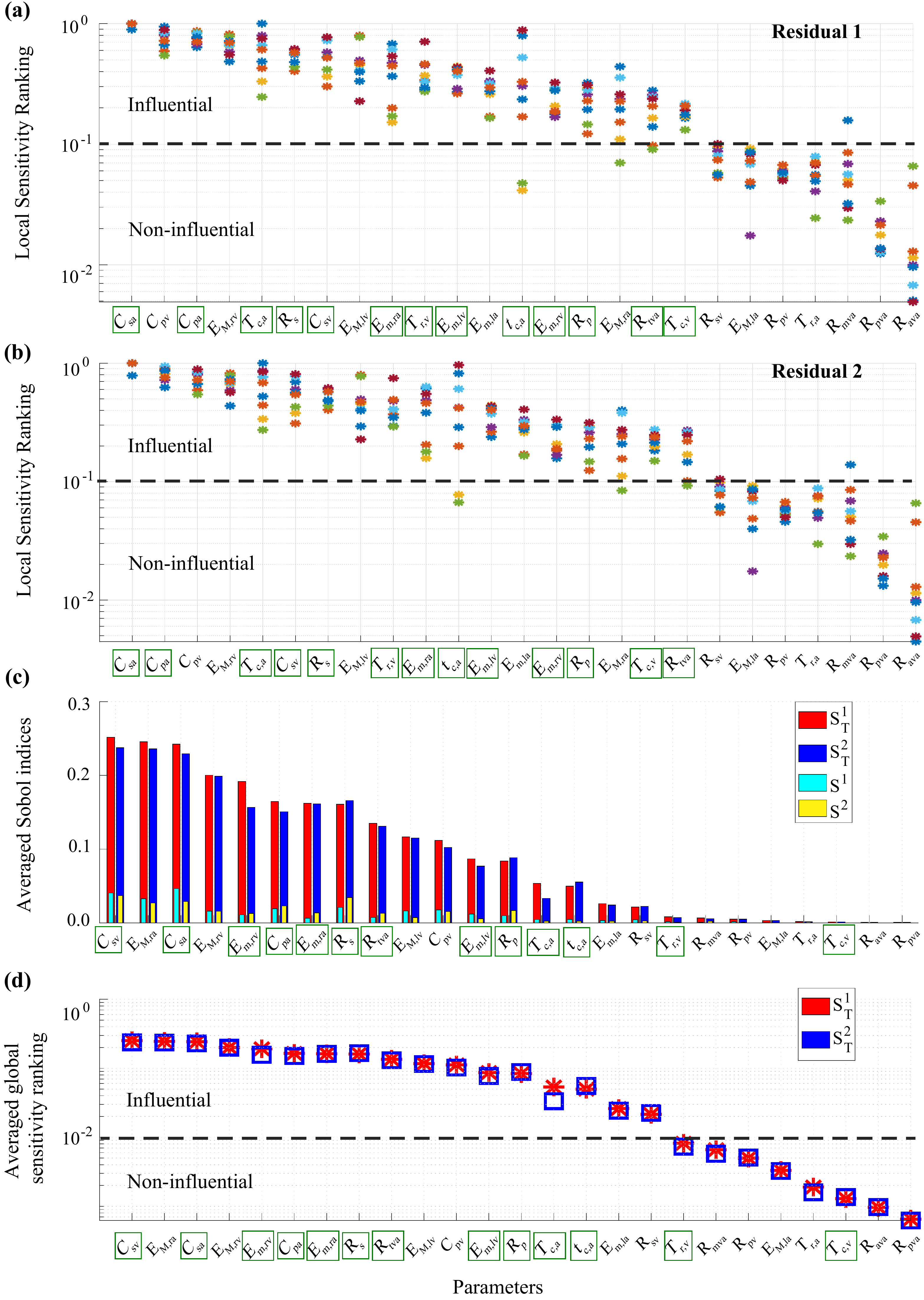}
    \caption{\textbf{Sensitivity analysis.} Parameter ranking based on a local sensitivity analysis using either $\bm{r}_1$ (a) or $\bm{r}_2$ (b). Each patient is plotted by a different color. The parameter order is based on median sensitivity across all patients. Average first order $(S_i)$ and total order $(S_{T_i})$ Sobol' indices using either $\bm{r}_1$ or $\bm{r}_2$ (c). Average parameter ranking based on $(S_{T_i})$ magnitude for either residual are shown in (d). The horizontal dashed lines separate influential (above) and non-influential (below) parameters.}
    \label{fig:Sensitivity_ranking}
\end{figure}

Figure \ref{fig:Sensitivity_ranking} (a-b) shows the patient-specific local sensitivity parameter ranking for $\mathbf{r}_1$ (static values only, panel (a)) and $\mathbf{r}_2$ (static and time-series data, panel (b)). Sensitivities are normalized by the largest magnitude for each patient and residual, and parameters are sorted based on their median ranking across all nine patients.

Parameters are ranked similarly for the two residual vectors; however, accounting for dynamic predictions makes the timing parameter $\tau_{c,a}$ more influential on $\mathbf{r}_2$. The most influential parameters for both residuals are $C_{sa},\, C_{pa},\, C_{pv},$ and $E_{M,rv}$. Seven of the nine patients display consistent parameter rankings for both residual vectors. Parameter $\tau_{c,a}$ is less influential for patients 3 and 5 than for the other patients. Overall, the local analysis shows that parameters [$R_{ava},\,R_{mva},\,R_{pva},\,R_{pv},\,R_{sv},\,E_{M,la}, \,T_{r,a}$] are non-influential for both residuals; parameters with sensitivities $\leq 10^{-1}$ are considered non-influential. The boundary separating influential and non-influential parameters is marked with vertical lines in figure \ref{fig:Sensitivity_ranking}.

For the global sensitivity analysis, $n=10^4$  samples are generated for each parameter using a Sobol' sequence. The average first order ($S_i$) and total ($S_{T_i}$) effects across all nine patients are shown in figure \ref{fig:Sensitivity_ranking} (c-d) for the cost functional $J(\theta)$ using residuals $\mathbf{r}_1$ and $\mathbf{r}_2$. Sobol' indices are similar across all patients, and the parameter ranking using the total Sobol' index agrees with the local results. A total index, $S_{T_i}$, near zero ($\leq 10^{-2}$) suggests that the corresponding parameter is non-influential. Results show that $S_{T_i}$ is $\leq0.005$ for parameters $R_{ava}$, $R_{pva}$, $R_{pv}$, $R_{mva}$, $E_{M,la}$, and $T_{r,a}$, consistent with the local sensitivity results, suggesting that these parameters can be fixed at their nominal values. The $S_{T_i}$ is also approximately zero for $T_{c,v}$ and $T_{r,v}$. Since the local sensitivity identifies $T_{c,v}$ and $T_{r,v}$ as influential, we include these in our subset selection procedure.

\subsection{Parameter subsets and inference}
Both SVD-QR and multi-start inference are used for parameter subset selection. The non-influential parameters, $\theta^{NI}=[R_{ava},\,R_{mva},\,R_{pva},\,R_{pv},\,E_{M,la}, \,T_{r,a}$] are fixed prior to SVD-QR. Previous studies \cite{Williams2019} found that the maximum and minimum elastance cannot be inferred simultaneously. Since the minimum elastance control both the amplitude and baseline elastance, this parameter contains more information and is, therefore, more important to infer. The study by Domogo and Ottesen \cite{Domogo2021} focused on left atrial dynamics using a 0D computational model. They found that changes in atrial volume were sensitive to maximal atrial compliance (i.e., minimal atrial elastance). This observation supports our exclusion of maximal elastance parameters in subset selection. Thus, the remaining maximal elastances, [$E_{M,ra},\, E_{M,rv},\, E_{M,lv}$], are also fixed prior to SVD-QR. We generate a subset for each residual, including parameters consistently identified by SVD-QR across all nine patients. Parameters that are inconsistent using SVD-QR are depicted in blue in tables S1 and S2 of the Supplemental Material.

\begin{figure}[ht!]
    \centering
    \includegraphics[width=0.65\linewidth]{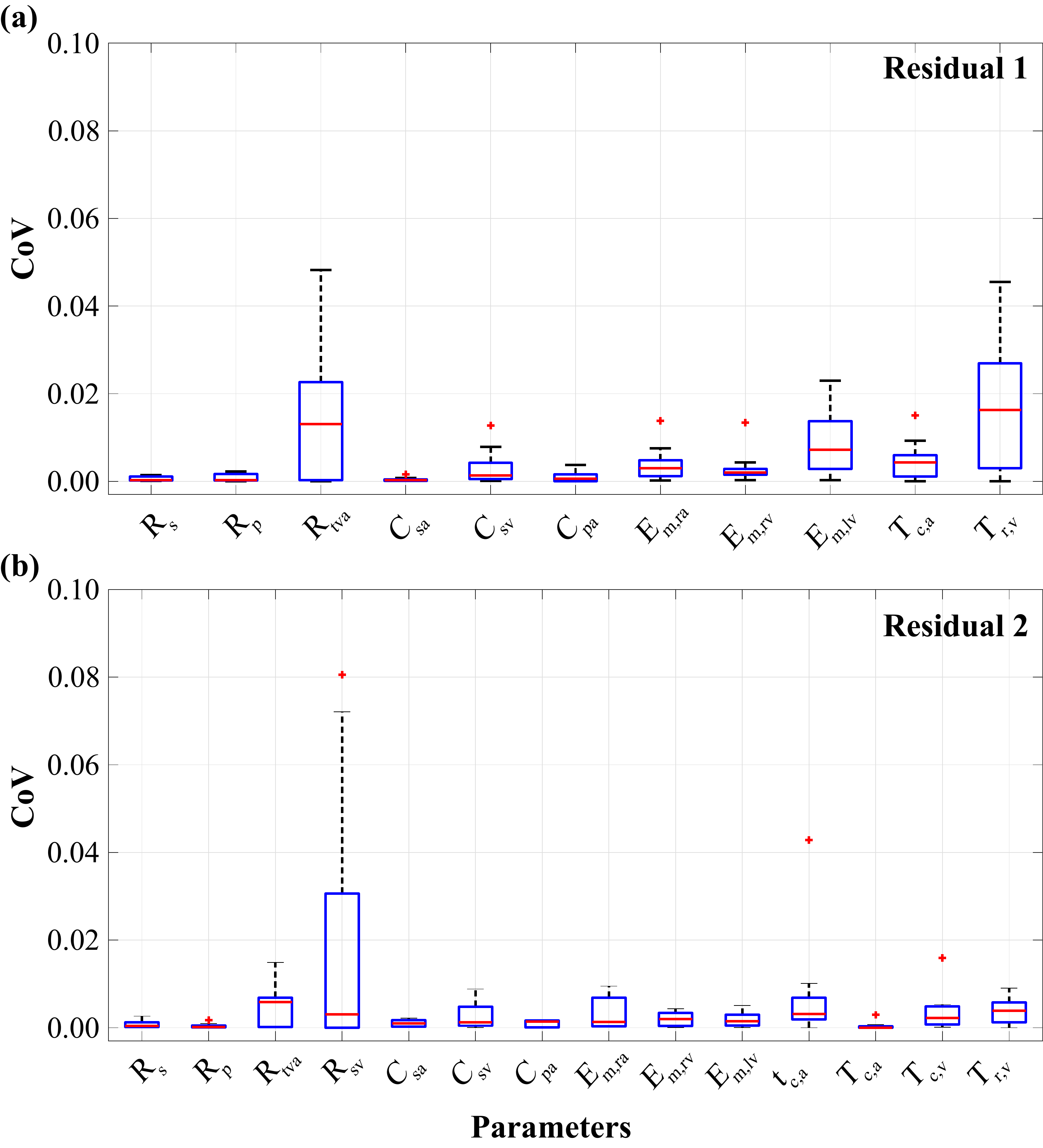}
    \caption{\textbf{Multistart Inference}. For each patient, each subset is tested for identifiability using 8 randomized starting guesses within $\pm 20\%$ of the nominal value. Coefficient of variance (CoV) for the final parameter sets using (a) $\bm{r}_1$ or (b) $\bm{r}_2$ provide a CoV below 10\%.}
    \label{fig:multistart}
\end{figure}

We run the multistart inference with these reduced SVD-QR subsets. For instances of multistart inference that have parameters with high CoV ($>0.10$) (purple in tables S1 and S2 in the Supplemental Material), the least influential parameter is removed from the subset and fixed at its nominal value. The final subsets used for each residual are
\begin{eqnarray}
    \bm{\theta}^{r_1} &=& \left[R_s, R_p, R_{tva}, C_{sa}, C_{sv}, C_{pa},E_{m,ra}, E_{m,rv}, E_{m,lv}, T_{c,a}, T_{r,v}\right],\\
 \bm{\theta}^{r_2} &=& \left[R_s, R_p, R_{tva}, R_{sv}, C_{sa}, C_{sv}, C_{pa},E_{m,ra}, E_{m,rv}, E_{m,lv},  \tau_{c,a}, T_{c,a}, T_{c,v}, T_{r,v}\right].
\end{eqnarray}
Figure \ref{fig:multistart} shows the CoV of the final subsets for $\mathbf{r}_1$ and $\mathbf{r}_2$. Table \ref{tab:ParR1} and \ref{tab:ParR2} list the estimated parameters using either $\mathbf{r}_1$ or $\mathbf{r}_2$. These optimal values reflect the optimization starting from the nominal guesses for each patient. We also calculate the 95\% parameter confidence intervals using eq. \eqref{eqn:25}.

\begin{table}[!htp]\centering
\caption{Estimated parameter values using $\mathbf{r_1}$ along with the $95\%$ confidence interval (depicted as $\hat{\theta_i}\pm 2\sigma_{\theta_i}$).}\label{tab:ParR1}
\begin{adjustbox}{width=1\textwidth}\small
\begin{tabular}{c|ccccccccc}\toprule
$\theta$ &\textbf{P1} &\textbf{P2} &\textbf{P3} &\textbf{P4} &\textbf{P5} &\textbf{P6} &\textbf{P7} &\textbf{P8} &\textbf{P9} \\\midrule\midrule
$R_s$ &0.82$\pm$0.14 &1.13$\pm$0.97 &1.2$\pm$0.65 &1.14$\pm$0.12 &1.08$\pm$0.82 &0.9$\pm$0.12 &0.86$\pm$0.16 &0.55$\pm$0.74 &1.24$\pm$0.12 \\
$R_p$ &0.5$\pm$0.11 &0.92$\pm$0.71 &0.77$\pm$0.42 &0.33$\pm$0.21 &0.59$\pm$0.87 &0.28$\pm$0.18 &0.36$\pm$0.22 &0.26$\pm$0.71 &0.33$\pm$0.18 \\
$R_{tva}$ &0.02$\pm$1.63 &0.12$\pm$3.95 &0.12$\pm$2.26 &0.07$\pm$0.57 &0.09$\pm$3.01 &0.01$\pm$3.29 &0.01$\pm$2.03 &0.03$\pm$1.57 &0.05$\pm$0.65 \\
$C_{sa}$ &2.11$\pm$0.02 &0.99$\pm$0.08 &1.45$\pm$0.08 &0.86$\pm$0.02 &0.98$\pm$0.09 &1.69$\pm$0.02 &2.18$\pm$0.02 &2.37$\pm$0.06 &0.67$\pm$0.02 \\
$C_{sv}$ &57.86$\pm$0.36 &22.36$\pm$0.7 &14.26$\pm$0.48 &45.68$\pm$0.45 &21.16$\pm$0.87 &34.67$\pm$0.18 &15.67$\pm$0.13 &13.4$\pm$0.41 &35.67$\pm$0.25 \\
$C_{pa}$ &1.33$\pm$0.01 &0.63$\pm$0.08 &0.82$\pm$0.06 &1.13$\pm$0.03 &0.82$\pm$0.1 &2.95$\pm$0.02 &1.84$\pm$0.02 &1.69$\pm$0.06 &0.98$\pm$0.03 \\
$E_{ra,m}$ &0.06$\pm$0.37 &0.23$\pm$0.71 &0.28$\pm$0.48 &0.1$\pm$0.44 &0.25$\pm$0.81 &0.1$\pm$0.18 &0.16$\pm$0.13 &0.26$\pm$0.39 &0.12$\pm$0.25 \\
$E_{rv,m}$ &0.03$\pm$0.62 &0.03$\pm$2.18 &0.02$\pm$3.51 &0.01$\pm$3.36 &0.07$\pm$0.83 &0.05$\pm$0.64 &0.09$\pm$0.1 &0.09$\pm$0.51 &0.03$\pm$0.56 \\
$E_{lv,m}$ &0.02$\pm$0.13 &0.03$\pm$2.66 &0.04$\pm$1.2 &0.07$\pm$0.15 &0.11$\pm$0.51 &0.05$\pm$0.35 &0.1$\pm$0.29 &0.13$\pm$0.76 &0.09$\pm$0.14 \\
$T_{c,a}$ &0.73$\pm$0.31 &0.74$\pm$0.54 &0.9$\pm$0.68 &0.9$\pm$0.23 &0.85$\pm$0.91 &0.83$\pm$0.85 &0.86$\pm$0.44 &0.67$\pm$0.94 &0.79$\pm$0.13 \\
$T_{r,v}$ &0.48$\pm$0.04 &0.52$\pm$2.22 &0.76$\pm$0.85 &0.5$\pm$0.48 &0.6$\pm$1.19 &0.56$\pm$0.63 &0.58$\pm$0.07 &0.54$\pm$0.62 &0.56$\pm$0.32 \\
\bottomrule
\end{tabular}
\end{adjustbox}
\end{table}    

\begin{table}[!htp] \centering
   \caption{Estimated parameter values using $\mathbf{r_2}$ along with the $95\%$ confidence interval (depicted as $\hat{\theta_i}\pm 2\sigma_{\theta_i}$).}\label{tab:ParR2}
   \footnotesize
    \centering
    \begin{adjustbox}{width=1\textwidth}\small
    \begin{tabular}{c|ccccccccc} \toprule
    \textbf{$\Theta$} & \textbf{P1} & \textbf{P2} & \textbf{P3} & \textbf{P4} & \textbf{P5} & \textbf{P6} & \textbf{P7} & \textbf{P8} & \textbf{P9} \\ \midrule\midrule
   $R_s$ &0.77$\pm$0.82 &1.14$\pm$1.31 &1.25$\pm$0.95 &1.12$\pm$0.32 &1.15$\pm$0.57 &0.9$\pm$0.2 &0.79$\pm$0.42 &0.58$\pm$0.28 &1.21$\pm$0.21 \\
$R_p$ &0.49$\pm$0.1 &0.9$\pm$0.14 &0.76$\pm$0.13 &0.34$\pm$0.1 &0.58$\pm$0.12 &0.28$\pm$0.07 &0.36$\pm$0.09 &0.26$\pm$0.05 &0.33$\pm$0.09 \\
$R_{tva}$ &0.03$\pm$0.36 &0.12$\pm$0.45 &0.12$\pm$0.3 &0.05$\pm$0.21 &0.09$\pm$0.31 &0.01$\pm$0.25 &0.03$\pm$0.23 &0.03$\pm$0.1 &0.05$\pm$0.19 \\
$R_{sv}$ &0.02$\pm$1.77 &0.02$\pm$2.77 &0.01$\pm$3.42 &0.01$\pm$1.84 &0.01$\pm$1.69 &0.01$\pm$1.07 &0.05$\pm$1.03 &0.01$\pm$0.58 &0.02$\pm$1.11 \\
$C_{sa}$ &2.04$\pm$0.12 &0.99$\pm$0.11 &1.5$\pm$0.14 &0.85$\pm$0.05 &1.01$\pm$0.06 &1.68$\pm$0.03 &2.12$\pm$0.06 &2.43$\pm$0.04 &0.66$\pm$0.03 \\
$C_{sv}$ &35.45$\pm$0.47 &23.24$\pm$0.23 &15.7$\pm$0.13 &42.86$\pm$0.21 &24.05$\pm$0.08 &33.5$\pm$0.1 &12.16$\pm$0.2 &14.42$\pm$0.02 &31.12$\pm$0.16 \\
$C_{pa}$ &1.3$\pm$0.02 &0.63$\pm$0.02 &0.89$\pm$0.02 &1.13$\pm$0.02 &0.83$\pm$0.02 &2.91$\pm$0.01 &1.79$\pm$0.01 &1.73$\pm$0.01 &0.95$\pm$0.01 \\
$E_{ra,m}$ &0.07$\pm$0.25 &0.22$\pm$0.25 &0.24$\pm$0.2 &0.11$\pm$0.16 &0.22$\pm$0.12 &0.1$\pm$0.07 &0.19$\pm$0.09 &0.23$\pm$0.06 &0.13$\pm$0.12 \\
$E_{rv,m}$ &0.03$\pm$0.38 &0.03$\pm$0.45 &0.02$\pm$0.97 &0.01$\pm$0.8 &0.08$\pm$0.08 &0.05$\pm$0.05 &0.1$\pm$0.06 &0.09$\pm$0.02 &0.03$\pm$0.16 \\
$E_{lv,m}$ &0.02$\pm$0.44 &0.03$\pm$0.61 &0.04$\pm$0.53 &0.06$\pm$0.22 &0.11$\pm$0.19 &0.05$\pm$0.08 &0.11$\pm$0.17 &0.14$\pm$0.09 &0.09$\pm$0.17 \\
$\tau_{c,a}$ &0.48$\pm$0.28 &0.43$\pm$0.43 &0.63$\pm$0.22 &0.69$\pm$0.1 &0.41$\pm$0.3 &0.6$\pm$0.09 &0.54$\pm$0.16 &0.44$\pm$0.08 &0.55$\pm$0.11 \\
$T_{c,a}$ &0.91$\pm$0.14 &0.74$\pm$0.17 &0.92$\pm$0.12 &0.9$\pm$0.08 &0.85$\pm$0.1 &0.87$\pm$0.05 &0.98$\pm$0.05 &0.69$\pm$0.03 &0.82$\pm$0.06 \\
$T_{c,v}$ &0.31$\pm$0.04 &0.27$\pm$0.05 &0.2$\pm$0.06 &0.29$\pm$0.03 &0.22$\pm$0.04 &0.34$\pm$0.02 &0.35$\pm$0.03 &0.24$\pm$0.02 &0.32$\pm$0.02 \\
$T_{r,v}$ &0.5$\pm$0.03 &0.51$\pm$0.04 &0.76$\pm$0.04 &0.62$\pm$0.03 &0.48$\pm$0.04 &0.56$\pm$0.02 &0.58$\pm$0.02 &0.51$\pm$0.01 &0.59$\pm$0.02 \\ \bottomrule
\end{tabular}
    \end{adjustbox}
\end{table}   
    
\begin{figure}[ht!]
    \centering
    \includegraphics[width=0.65\linewidth]{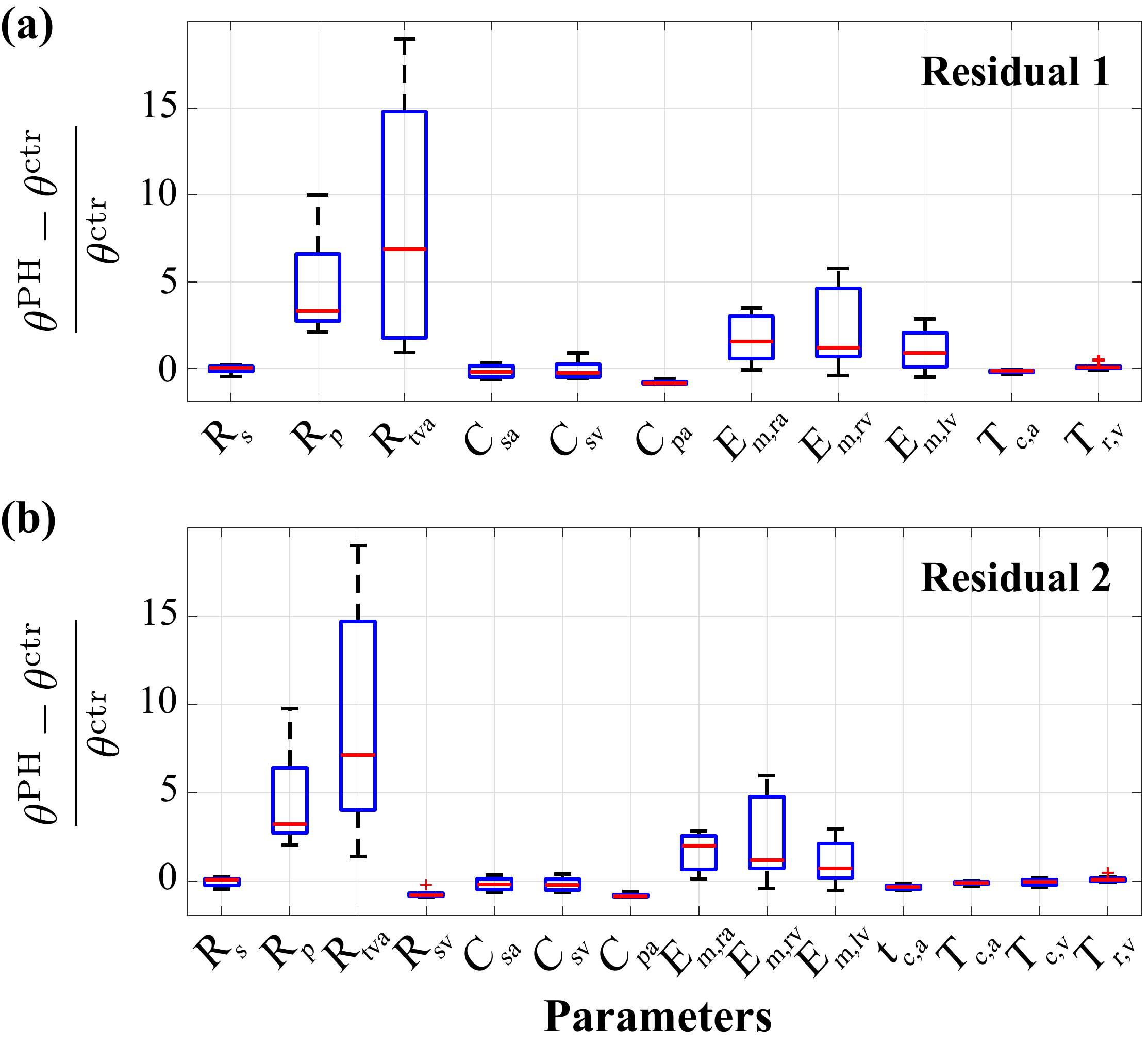}
    \caption{\textbf{Changes in parameters due to PH.} Box and whisker plots showing quantiles and outliers for the estimated parameters. Results show the relative difference from the normotensive predictions.}
    \label{fig:RelDiff}
\end{figure}

\begin{figure}[ht!]
    \centering
    \includegraphics[width=0.65\linewidth]{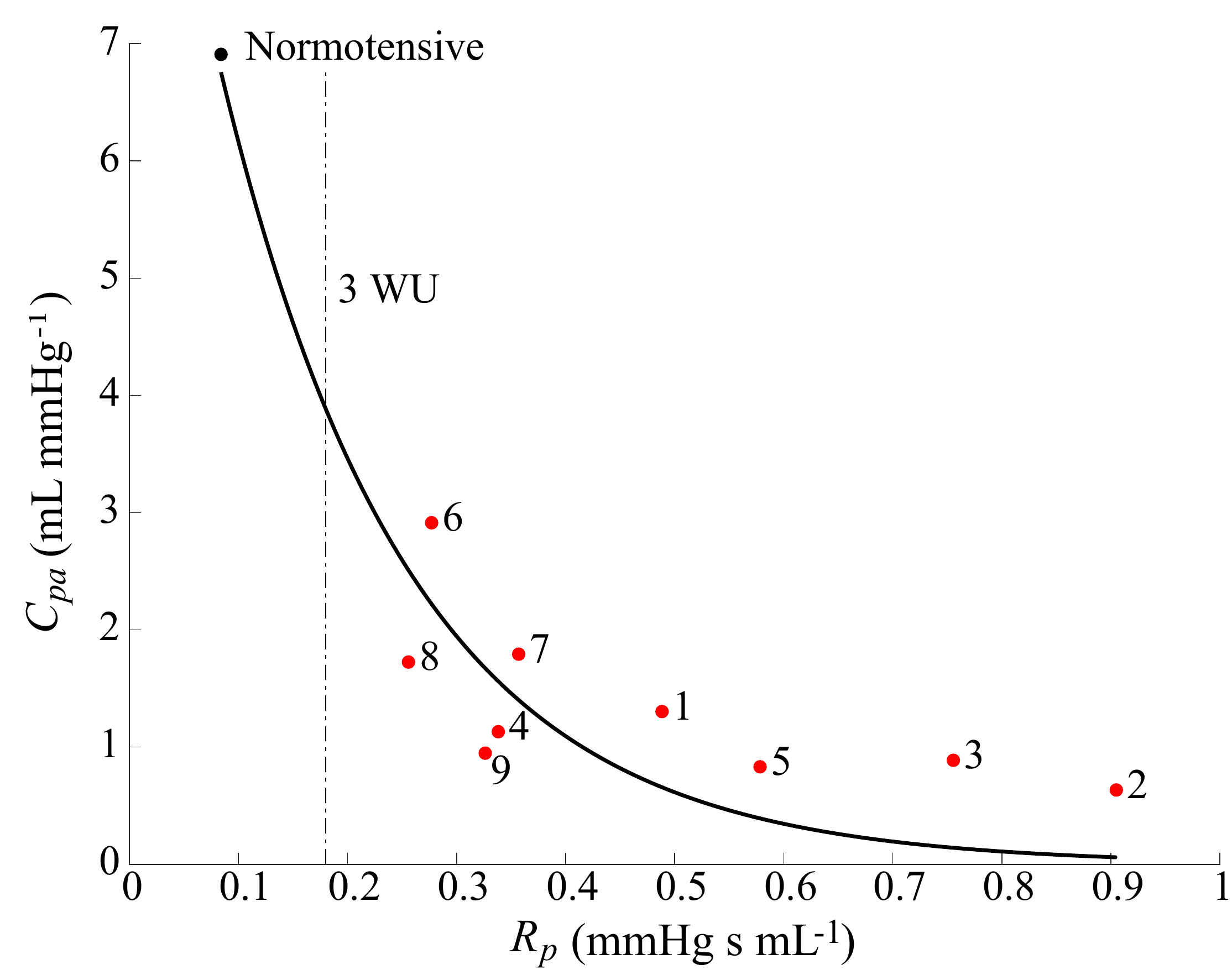}
    \caption{\textbf{Hyperbolic $R_{p}$-$C_{pa}$ relationship}. Optimal values of $R_{p}$ and $C_{pa}$ for the normotensive (black) and PH (red) patients. The best-fit curve is given by $C_{pa} = 0.6518/(0.1005 +R_p)$, and is similar to previous findings using isolated Windkessel models \cite{Tedford2012}. A PVR $\geq$ 3 Wood units (3 WU = 0.18 mmHg s mL$^-1$) is considered in the PH range (dashed line).}
    \label{fig:RC}
\end{figure}

We display the relative change between estimated PH parameters and normotensive parameters in figure \ref{fig:RelDiff} as box-and-whisker plots to understand how parameters change with PH. Note that estimated parameters shared between $\bm{\theta}^{r_1}$ and $\bm{\theta}^{r_2}$ are nearly identical even with additional parameters in $\bm{\theta}^{r_2}$. Parameters $R_p$, $R_{tva}$, $E_{m,ra}$, $E_{m,rv}$, and $E_{m,lv}$ are consistently elevated in all PH patients. The normotensive value of $R_{tva}$ is substantially smaller than the PH patients, which explains the larger relative change compared to other parameters in the subset. The timing parameters for the heart chambers, compartment compliances, and systemic resistances $R_s$ and $R_{sv}$ remain relatively close to normotensive values. The $R_p$-$C_{pa}$ (RC) relationship was also determined from the inferred parameters. As shown in figure \ref{fig:RC}, there is a clear inverse relationship between $R_p$ and $C_{pa}$ with the curve of best fit being $C_{pa} = 0.6518/(0.1005 +R_p)$, $R^2 = 0.77$, and constant RC time $R_p\cdot C_{pa} = 0.55\pm0.15$s.

\subsection{Model forecasts and uncertainty}

\begin{figure}[ht!]
    \centering
    \includegraphics[width=0.8\linewidth]{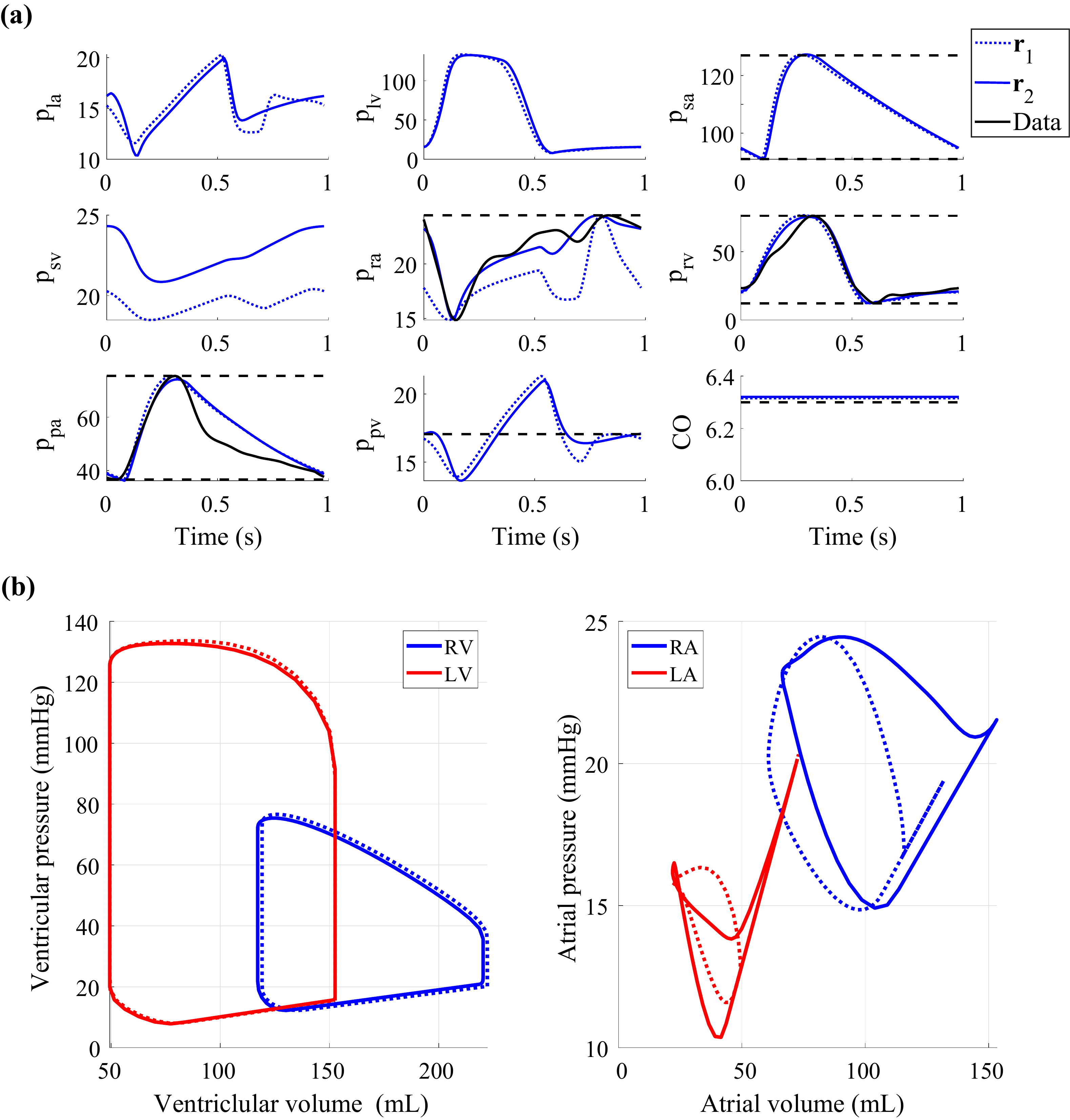}
    \caption{\textit{Optimal model predictions.} (a) Optimal model fits for pressure, $p_i$ (mmHg) and cardiac output, CO (L/min) using either $\bm{r}_1$ (dotted line) or $\bm{r}_2$ (solid line) compared to the data for patient 7. (b) simulated pressure-volume loops in the ventricles and atria using residual 1 (dotted line) and residual 2 (solid line) for the dataset from patient 7.}
    \label{fig:optP7}
\end{figure}

Post-inference predictions of pressure and CO using either $\bm{r}_1$ or $\bm{r}_2$  are depicted in figure \ref{fig:optP7}(a) along with the measured data from patient 7. Predictions for all PH patients are included in the Supplemental Material. Both $\bm{r}_1$ and $\bm{r}_2$ inference procedures are able to match the static data well. Using $\bm{r}_2$ minimizes the mismatch between the dynamic model outputs and the time-series data. Predictions of RA dynamics improve drastically when including time-series data. In contrast, RV and PA predictions improve only marginally. For five patients, CO predictions are only slightly worse when matched using $\bm{r}_2$ vs. $\bm{r}_1$. However, maximum and minimum pressure values still match the data well.

\begin{figure}[ht!]
    \centering
    \includegraphics[width=0.85\linewidth]{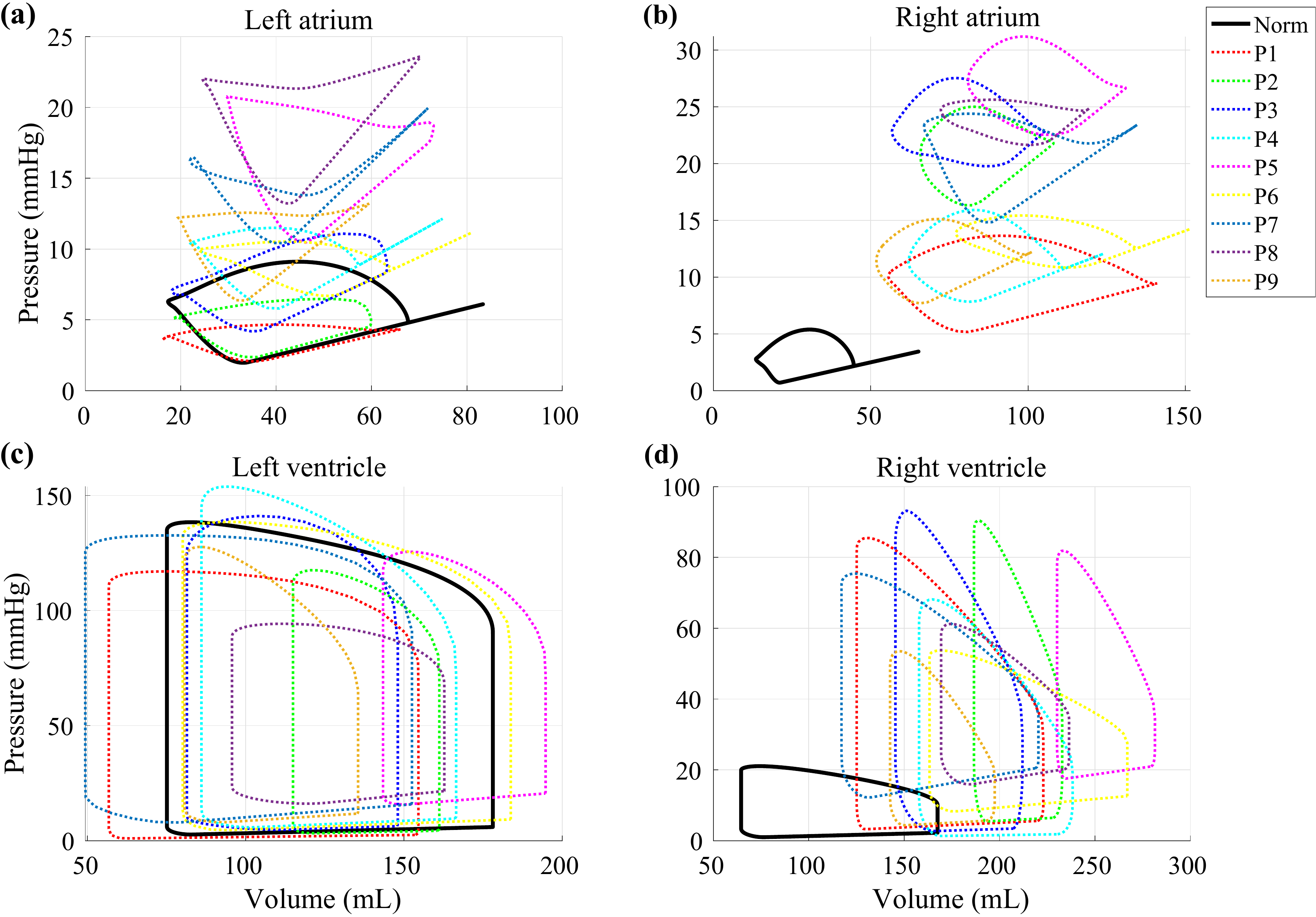}
    \caption{\textbf{Simulated pressure-volume loops.} Pressure-volume loops in the normotesive (norm) and all nine PH patients are contrasted. Model predictions include (a) left atrial, (b), right atrial, (c) left ventricular, and (d) right ventricular pressure-volume loops.}
    \label{fig:PV_compare}
\end{figure}

A benefit of computational models is that essential but unmeasurable outcomes, such as PV loops, can be predicted. We contrast PV loops from all four heart chambers for the normotensive subject to the nine PH patients (using estimated parameters from $\mathbf{r}_2$) in figure \ref{fig:PV_compare}. Except for patients 1 and 2, all PH patients have increased left atrial pressure. In contrast, RA PV loops display elevated volumes and pressures relative to the normotensive simulation for all patients. The RV and LV PV loops have similar shapes, yet the RV PV loops in the PH group have a more drastic rise in pressure during isovolumic contraction compared to the normotensive results.

We calculate SW for all four heart chambers by integrating simulated pressure with respect to volume. These results and other model outcomes, including the resistance and compliance ratios, $R_p/R_s$ and $C_{pa}/C_{sa}$, and the pulsatility index PI, are shown in Table \ref{tab:outcomes}. Left atrial SW is lower in PH for all but patients 5 and 8, and RA SW is higher in all PH patients relative to the normotensive value. LV SW is lower in all CTEPH patients (3, 4, 5, and 9) and two PAH patients (2 and 8), while RV SW is increased in all nine PH patients. In general, there is a drastic increase in $R_p/R_s$ and decrease in $C_{pa}/C_{sa}$ in PH relative to normotensive conditions. The PI decreased in PH except in patient 1.

\begin{table}[!htp]
\caption{Model outcomes from normotensive and PH simulations.}\label{tab:outcomes}
\begin{center}
\begin{tabular}{c|ccccccccc}
\toprule
& \multicolumn{4}{c}{SW} & & & \\
\cline{2-5}
\textbf{Patient} &\textbf{LA} &\textbf{LV} &\textbf{RA} &\textbf{RV} &$\mathbf{R_p/R_s}$ &$\mathbf{C_{pa}/C_{sa}}$ &\textbf{PI} \\ \midrule\midrule
Norm & 0.031 & 1.676 & 0.013 & 0.223 & 0.08 & 3.84 &  4.25 \\
1            & 0.010 & 1.556 & 0.064 & 0.882 & 0.63 & 0.64 & 5.37 \\
2            & 0.020 & 0.868 & 0.041 & 0.532 & 0.79 & 0.64 & 2.40 \\
3            & 0.023 & 1.150 & 0.034 & 0.618 & 0.60 & 0.59 & 2.52 \\
4            & 0.018 & 1.502 & 0.039 & 0.590 & 0.30 & 1.33 & 3.94 \\
5            & 0.038 & 0.779 & 0.043 & 0.368 & 0.50 & 0.83 & 1.63 \\
6            & 0.012 & 1.728 & 0.024 & 0.488 & 0.31 & 1.74 & 1.87 \\
7            & 0.009 & 1.618 & 0.042 & 0.640 & 0.45 & 0.84 & 1.76 \\
8            & 0.038 & 0.892 & 0.022 & 0.423 & 0.44 & 0.71 & 1.07 \\
9            & 0.021 & 0.888 & 0.035 & 0.324 & 0.27 & 1.44 & 2.85 \\ \bottomrule
\end{tabular}
\vspace{2mm}
\end{center}

{\footnotesize Indices include stroke work (SW, Joule) in all four heart chambers, resistance ratios (dimensionless), compliance ratios (dimensionless), and pulsatility index (PI, dimensionless) calculated after estimating parameters using $\bm{r}_2$. LA – left atrium, LV – left ventricle, RA – right atrium, RV – right ventricle.}
\end{table}

Parameters confidence intervals are provided in table \ref{tab:ParR2}. Model confidence and prediction intervals for patient 7 are shown in figure \ref{fig:UQP7} (see the Supplemental Material for results from all nine patients) using either residual vector. The confidence and prediction intervals show uncertainty in mean pulmonary venous pressure (matched to PAWP data), CO, and maximum and minimum pressures in the systemic arteries, RA, RV, and PA. The confidence intervals for RV and PA are smaller than the RA and are attributed to the larger mismatch between RA data and model simulations. Adding dynamic data in $r_2$ increases the magnitude of the sum of squared residuals, thus increasing the prediction intervals in figure\ref{fig:UQP7}b. Note that the PA, RA, and RV data nearly all fall within the 95\% prediction intervals shown in figure\ref{fig:UQP7}b.

\begin{figure}[ht!]
    \centering
    \includegraphics[width=0.9\linewidth]{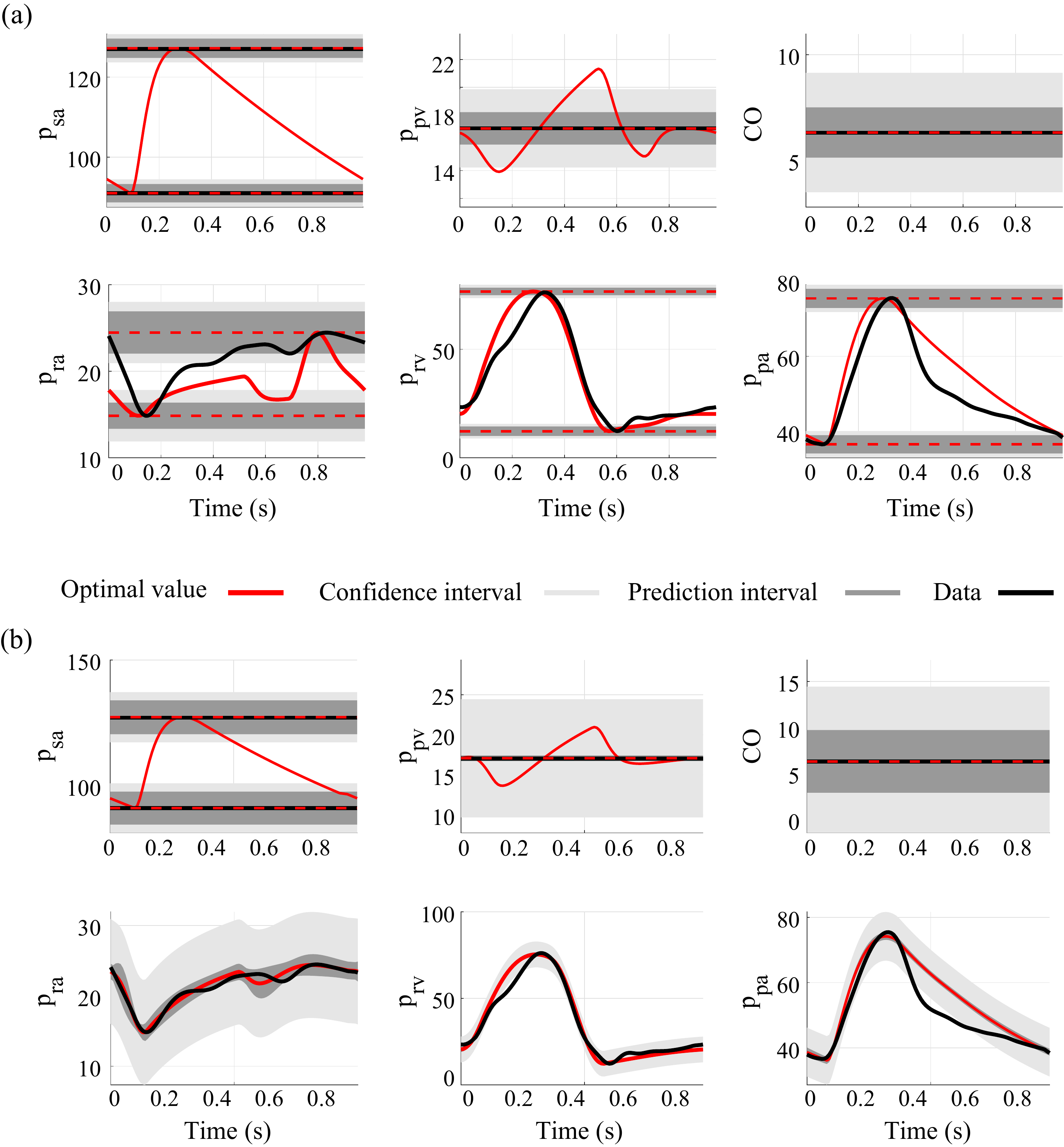}
    \caption{\textbf{Output uncertainty.} Uncertainty in the model outputs for pressure, $p_i$ (mmHg) and cardiac output, CO (L/min) using either $\bm{r}_1$ (a) or $\bm{r}_2$ (b) for the quantity of interest.}
    \label{fig:UQP7}
\end{figure}

\section{Discussion}

Electronic health records typically include RHC blood pressure measurements, estimates of cardiac output, and systolic and diastolic blood pressure cuff measurements in the systemic circulation. Traditionally, static pressures (e.g., systolic \& diastolic) are recorded, though the RHC also generates blood pressure waveforms. Our goal is to examine if additional waveform data improve model calibration and, therefore, characterization of PH and its phenotypes. We use a systems-level cardiovascular model to characterize patient-specific changes due to PH. We use a combination of sensitivity analyses, subset selection, and multi-start inference to determine informative and identifiable parameter subsets and estimate these parameters to patient RHC data. Results show that the proposed model captures the hallmarks of PH both with and without waveform data. We find increased RA, RV, and PA pressures, elevated PVR, and reduced pulmonary arterial compliance in all PH patients. Finally, we show that additional waveform data are essential in quantifying RA reservoir and pump function. Overall, our results show that systems-level models can capture patient-specific PH dynamics and parallel the current clinical understanding of the disease.

\subsection{Sensitivity analyses}

Sensitivity analysis is crucial for determining which parameters influence the model output. Our model has 25 parameters, yet limited data and the structure of the model make inferring all the parameters infeasible. We use local and global sensitivity analyses on two residual vectors: one comparing static outputs and another static and dynamics outputs. Both methods consistently identify 16 influential and six non-influential parameters, independent of the technique and residual. Three parameters, $[R_{sv},\,T_{c,v}, \,T_{r,v}]$, are excluded from the sets as they are not consistently influential across the two techniques. The influential parameters are candidates to be inferred, while the non-influential parameters will be kept fixed at their nominal value.

The pulmonary valve resistance ($R_{pva}$) is non-influential; this parameter is directly associated with the coupling between the RV and PA. However, none of the PH patients in this study have a history of pulmonary valve stenosis. Thus it is reasonable to keep this parameter fixed at its nominal value. The pulmonary venous ($R_{pv}$) and mitral valve ($R_{mva}$) resistances are also not influential. Since we do not have left heart data, the residuals do not include left heart quantities, and therefore we expect these to be non-influential. This finding agrees with previous studies \cite{Marquis2018,Ellwein2008,Harrod2021} that fix the valve resistances.

Both local and global analysis techniques are essential as they each highlight model features. Global sensitivities identify influential parameters over the physiological parameter range, while local sensitivities are evaluated at known values. Global sensitivity analysis sample parameters over the physiological range, but due to nonlinear model behavior, this could include combinations that generate an non-physiological output. Yet, the local analysis only provides a snapshot of the sensitivities; again, since the model is nonlinear, the parameter influence may change if a parameter is changed, i.e., a parameter influential before optimization could be non-influential after optimization. For example (see figure\ref{fig:Sensitivity_ranking}), the atrial timing parameter $\tau_{c,a}$ is less influential for patients 3 and 5 than for the other PH patients, and $E_{M,la}$ is less influential for patient 4. These results agree with Marquis et al. \cite{Marquis2018}, where LV elastance and systolic timing parameters varied across each rat. Global sensitivity analysis cannot identify these discrepancies, as it integrates the sensitivity over the physiological parameter space.

Finally, while influential parameters are consistent between methods, individual parameters may have a different ranking. As shown in Figure\ref{fig:Sensitivity_ranking}, the maximal atrial elastance $E_{ra,M}$ is the second most influential parameter in the global analysis, whereas the local analysis ranks the parameter significantly lower. This can be attributed to interactions between $E_{M,ra}$ and $E_{ra,m}$, which account for the RA reservoir and pump function. Small changes in $E_{ra,m}$ drastically affect maximum and minimum pressure values while changes in $E_{ra,M}$ only affect the model output when $E_{ra,M} \gg E_{ra,m}$. Thought the ranking of $E_{M,ra}$ differs, $E_{ra,m}$ is always influential. 

Deficiencies in RA reservoir and contractile function are strong predictors of mortality in PH \cite{Sudar2018}. RA filling during ventricular diastole is dictated by systemic venous dynamics and tricuspid valve integrity. In the model, RA systolic and diastolic pressures are determined by minimum elastance $E_{ra,m}$, which is always influential. The tricuspid valve resistance $R_{tva}$ forms the interface for RA-RV interactions. Hence, this parameter influences the relationship between the two heart chambers throughout the cardiac cycle. The high sensitivity of RA predictions to these parameters mimics the current physiological understanding of altered RA function in PH \cite{Sudar2018}.

Two of the three parameters characterized differently between the local and global methods are timing parameters dictating contraction and relaxation of the heart. The timing of heart contraction and relaxation are well approximated from dynamic pressure data. Hence, the uncertainty in these parameters (i.e., the bounds for global sensitivity sampling) is substantially smaller ($\pm 10-15\%$) than other model parameter uncertainty ($\pm 400\%$). This contributes to why the Sobol' indices are smaller than the local analysis. Since our nominal timing parameter values are well informed, the local analysis is more relevant and used to determine timing parameter influence. 

The final parameter with varying influence is $R_{sv}$, the systemic venous resistance. This parameter impacts central venous pressure and RA filling. As we discuss later, while at the border between influential and non-influential, the parameter is essential to predict atrial dynamics.

\subsection{Parameter inference and subset selection}

We fix non-influential parameters at their nominal values; however, this does not guarantee that the parameter subset is practically identifiable \cite{Miao2011,Harrod2021}. We combine SVD-QR subset selection and multistart parameter inference to determine an identifiable parameter subset. SVD-QR methods reduce the number of parameters \cite{Pope2009}, and multistart inference tests if solutions to the inverse problem are unique. For each patient, our results provide consistent parameter estimates across both residuals.
Results reveal that the model with static data has 11 identifiable parameters, while the model with static and dynamic data has 14 identifiable parameters. An important observation is that the identifiable parameter subsets are subsets of each other, i.e., $\bm{\theta}^{r_1}\subset \bm{\theta}^{r_2}$. These results demonstrate that the patient-specific model is robust.

Our finding that sensitivity analysis alone is inadequate to determine identifiable parameters agrees with results reported in the literature. For example, Schiavazzi et al. \cite{Schiavazzi2017} reported that sensitivity analyses does not guarantee unique parameter estimates. The authors use multistart inference to interrogate parameter identifiability and reduce their parameter subset. We use a similar technique. A CoV cutoff of 10\%, shown in figure \ref{fig:multistart}, ensures that parameter estimates are robust to simulations with 20\% uncertainty in initial guesses.

As shown on Figure \ref{fig:RelDiff}, identifiable parameters $R_p$, $R_{tva}$, $E_{m,ra}$, $E_{m,rv}$, and $E_{m,lv}$ are elevated in PH. The parameters $R_p$ and $R_{tva}$ have the largest relative increase. PVR is a known biomarker of PH disease severity, it is elevated in both PAH and CTEPH \cite{Humbert2010,Vonk2017}. The increase in minimum elastance in the RA and RV indicates chamber stiffening, as reported in PH \cite{Tello2019}. An elevated end-diastolic elastance, $E_{m,rv}$, is negatively correlated with RA reservoir, passive, and active strain \cite{Tello2019}, suggesting that RA and RV functions deteriorate during PH progression. We also observe a slight elevation in minimal LV elastance $E_{m,lv}$, correlating with impaired LV function due to rightward septal bulging \cite{Palau-Caballero2017}. Another important disease biomarker is pulmonary arterial compliance $C_{pa}$, which measures arterial distensibility. Figure \ref{fig:RelDiff} shows a relative decrease in $C_{pa}$ with PH, which consistent with literature \cite{Gelzinis2022}, reflects the stiffening of the proximal pulmonary arteries due to constitutive changes (e.g., collagen accumulation) \cite{Guigui2020}.

Several studies \cite{Gelzinis2022,Vonk2017,Tedford2012,Assad2016,Lankhaar2006} have emphasized the inverse relationship between $R_p$ and $C_{pa}$ in the pulmonary circulation, often referred to as RC-time, $\tau = R_p C_{pa}$. Tedford et al. \cite{Tedford2012} report an inverse-hyperbolic relationship from analysis of data from 1,009 patients with PH and normal pulmonary capillary wedge pressure with best-fit curve $C_{pa} = 0.564/(0.047+R_{p})$ and RC time $\tau = 0.48\pm0.17$. Similarly, the retrospective study by Assad et al. \cite{Assad2016} found that the RC time is $\tau = 0.7\pm0.34$ in PAH patients (n=593) with a best-fit curve $C_{pa} = 0.775/(0.045+R_p)$. They also noted that the inverse-hyperbolic RC-time relationship is nearly identical for PAH and group 2 PH patients. Figure \ref{fig:RC} shows this relationship from our patient cohort. The best fit curve $C_{pa} = 0.6518/(0.1105+R_p)$ and constant RC time $\tau = 0.55\pm0.15$ are consistent with results from these studies \cite{Tedford2012,Assad2016}. Our results were obtained from analysis of a closed-loop model, whereas the original RC times are computed using an isolated Windkesel model. This suggests that our systems-level model reproduces key features across large PH cohorts.

The parameters in the static and dynamic residuals, including the systemic venous resistance controlling flow from the systemic veins to the RA, significantly affect RA filling. PH patients have a small reduction in $R_{sv}$ relative to the normotensive patients,  increasing systemic venous inflow and diastolic RA filling. Growing evidence suggests that RA function is impaired during PH, though little is known about how RA-RV coupling is altered during disease progression \cite{Fayyaz2018,Sudar2018}. Using dynamic RA data for model calibration may provide new insight into the mechanisms of RA contractile and reservoir deterioration with RV dysfunction. Changes in RA contractile timing can only be observed with dynamic pressure data. Other parameters only in the dynamic residual include $T_{c,v}, T_{c,a}$, and $\tau_{c,a}$. These parameters are all associated with the timing of heart function, i.e., the generation of the waveforms. Alenezi et al. \cite{Sudar2018} studied RA strain across across 67 PAH subjects using speckle-tracking imaging. The study found that RA dysfunction was an independent predictor of mortality, and that RA strain rate (which is time dependent) correlate with PAH severity.  Future investigations using modeling with RA pressure and strain data may reveal additional indicators of RA dysfunction and PAH severity.

As shown in figure \ref{fig:optP7}, including more data in the parameter inference procedure not only increases the number of identifiable parameters but also changes model predictions and inferred parameter values. Both residuals account for systolic, diastolic, and mean values, which are well matched by the model across all patients. Dynamic PA and RV predictions are unchanged between $\mathbf{r}_1$ and $\mathbf{r}_2$. This is attributed to good nominal estimates of the ventricular timing parameters $T_{c,v}$ and $T_{r,v}$, i.e., the optimized values are close to nominal values. In contrast, there is a significant change in simulated RA dynamics when calibrating the model to dynamic pressure data. The intricate dynamics of atrial filling and contraction make it difficult to identify the RA timing parameters from pressure data visually. The PV loops in figure \ref{fig:optP7} show large changes in atrial dynamics when comparing $\mathbf{r}_1$ to $\mathbf{r}_2$. The study by Domogo and Ottesen \cite{Domogo2021} studied left atrial dynamics using a systems-level model. Their model has a more sophisticated atrioventricular coupling, but the authors noted that an elastance model can capture dynamic atrial data. The time-varying dynamics of the atria are more complex, demonstrating the need for dynamic rather than static data for model calibration. The RA is gaining traction as a biomarker for PH severity \cite{Sudar2018,Tello2019}. Hence our ability to calibrate RA dynamics may provide further insight into the progression of RA-RV-PA dysfunction in PH.

In the absence of volume data, we included additional volume constraints in our inference procedure. It is well established that both PAH and CTEPH cause increased RV myocardial remodeling, including wall thickening and dilatation \cite{Tello2019}. Penalizing the inference procedure to ensure BSA-indexed blood volumes in all cardiac chambers constrains the model forecasts to volumes seen in clinical studies \cite{Tello2019}. The addition of constraints leads to increased RA filling volumes and pressure magnitudes, as noted by Tello et al. \cite{Tello2019}. Moreover, as shown in \ref{fig:PV_compare}, the RV PV loop has a rightward shift but is comparable in shape to its LV counterpart. This shift is known to occur in PH \cite{Tabima2017}, increasing RV end-systolic elastance. While not modeled explicitly, our results show a reduction in LV PV loop area and SW due to RV dysfunction. A recent study by Jayasekera et al. \cite{Jayasekera2022} reported significant decreases in LV strain and prominent LV mechanical dyssynchrony in a cohort of patients with PAH. 

We predicted several outcomes using our model simulations, including Cardiac SW, a known indicator of cardiac oxygen consumption and overall cardiomyocyte function. Clinically, SW is calculated as the product of stroke volume and mean arterial pressure; using the model, SW is calculated more accurately by determining the area inside the PV loop. Both left and right heart SW, listed in table \ref{tab:outcomes}, change in PH. In general, LV SW decreases while right heart SW increases in PH. These findings agree with the retrospective clinical analysis by Chemla et al. \cite{Chemla2013}, who found that RV SW is doubled in PH. Increased RV SW is linked to severe pediatric PAH in a study by Yang et al., who also use a compartment model to generate PV loops. Without volume data, our model can provide these indicators of disease severity, making them clinically relevant.

\subsection{Uncertainty quantification}
We efficiently determined both parameter and output uncertainty using frequentist analyses. This study only infers identifiable parameters. Parameters that are more influential have narrower confidence intervals compared to less influential parameters (see table \ref{tab:ParR2}). A consequence of narrow parameter bounds is that the model confidence and prediction intervals that are sensitive to these influential parameters contain the measured data remarkably well for both residuals.
 
Output uncertainty is compared in figure \ref{fig:UQP7} for the two residuals $\mathbf{r}_1$ or $\mathbf{r}_2$. Model outputs computed using $\mathbf{r}_1$ have relatively small uncertainty for static targets. For $\mathbf{r}_2$, including both static and dynamic data, the uncertainty increases significantly, likely due to the increased complexity of the inverse problem. The least squares error is significantly higher, and even though the model does an excellent job fitting data, there are parts of the waveform that the simple lumped model used here cannot reproduce. However, we gain information about the dynamic output uncertainty in dynamic RA, RV, and PA predictions using $\mathbf{r}_2$. This better quantifies the expected beat-to-beat variation we would expect to see on continuous RHC monitoring. In general, a more liberal estimate of uncertainty as show from $\mathbf{r}_2$ reduces the chance of having a biased prediction due to a single heart beat of data.

Other groups have performed uncertainty quantification on cardiovascular models. The study by Harrod et al. \cite{Harrod2021} investigated PA pressure uncertainty using Markov chain Monte Carlo sampling. Their study focuses on uncertainties in model outputs using a normotensive parameter set, whereas our work explores output uncertainty using parameters indicative of PH. To our knowledge, this is the first study to consider output uncertainty in a systems-level cardiovascular model of PH. Several authors have performed uncertainty quantification using one-dimensional \cite{Colebank2019,Paun2020} or three-dimensional \cite{Schiavazzi2016} fluid dynamics models, which are fundamentally different than the systems-level model used here. Colebank et al. \cite{Colebank2019} found that uncertainty bounds around PA pressures were nearly identical between frequentist or Bayesian methods. The study also compared uncertainty across normotensive and hypoxia-induced PH mice. It showed larger uncertainty in the normotensive mice due to a larger discrepancy in the model fit to data. We see a similar trend in our results, with larger uncertainty typically attributed to patients with more complex RA dynamics (see Supplemental Material). Our 0D model cannot capture the dynamics of wave reflections suitable for a one-dimensional hemodynamics model. Yet, it does capture the global diastolic decay in PA pressure, as shown in figure \ref{fig:UQP7}. We match the model to RV dynamics exceptionally well; note the narrow confidence and prediction intervals in figure \ref{fig:UQP7}. The study by Yang et al. \cite{Yang2018} captured RV mechanics in PH using an simplified, open loop model. We show that a more complex model accounting for the systemic circulation and left heart can still predict RV dynamics with high accuracy.

\subsection{Limitations}
This study has several limitations. Our model accounts for LV and RV dynamics without including interventricular interaction through the septal wall. Several studies have included this mechanism in the modeling framework \cite{Palau-Caballero2017,Colunga2020}, which is important for understanding RV affects on LV function. Adding this model component provides a next step in understanding biventricular function during PH progression \cite{Jayasekera2022}. We use data from 9 patients, 4 of which have CTEPH while the other 5 are PAH. We do not have a sufficiently large sample size to deduce differences in PH phenotypes, though recent studies have found differences in the biomechanics of the two subgroups \cite{Raza2022}. Our inference procedure enforces cardiac volumes that match previously recorded BSA-indexed values; additional volume data (e.g., from a conductance catheter) would better inform the model calibration. Yet these were not available for the patients studied. Lastly, it is well established that PH disproportionately affects women, with sex differences being a significant area of attention in the PH community \cite{Cheng2022}. Combining a larger, more diverse patient cohort with the parameter inference performed here may elucidate sex-dependent differences in RA, RV, and PA parameters. Our study is a proof of concept that patient-specific models can be constructed from RHC data, laying the foundation for future studies on a larger population of patients.

\section{Conclusions}
This study uses a 0D, systems-level hemodynamics model to predict changes in cardiovascular parameters due to PH. We utilize sensitivity analyses and subset selection techniques to deduce the best parameter subsets for two residuals: one with static data and one with additional dynamic RA, RV, and PA pressure waveforms. Our results show that adding time-series waveform data allows for additional parameters $R_{sv},\,\tau_{c,a},\,T_{c,v}$ to be estimated without altering estimates in the static-only residual. These additional parameters better describe RA pump and reservoir function, which has been the focus of recent attention in the PH community \cite{Sudar2018}. Overall, model outcomes are consistent with the physiological understanding of the disease; PH induces increased PVR, decreased pulmonary arterial compliance, and elevated minimum RA and RV elastance, leading to increased mPAP. While the uncertainty in model predictions is smaller for the static residual, adding time-series data provides useful insight into uncertainty in dynamic predictions. Our study provides evidence that systems-level models can be tuned to fit PH data. The model can predict the right atrial function by adding static and dynamic data, which is important for differentiating PH subtypes. The framework devised here may be able to explain the mechanisms contributing to abnormal RA, RV, and PA function in PH.

\section*{Citation Diversity Statement}
In agreement with the editorial from the Biomedical Engineering Society (BMES) \cite{Rowson2021} on biases in citation practices, we have analyzed the gender and race of our bibliography. This is done manually, though automatic probabilistic tools exist (e.g.,  \url{https://zenodo.org/record/4104748#.YvVXpnbMI2z}). We recognize existing race and gender biases in citation practices and promote the use of diversity statements encouraging fair gender and racial author inclusion.

Our references, including those in the Supplemental Material, contain 15.15\% woman(first)/woman(last), 13.64\% man/woman, 16.67\% woman/man, and 54.55\% man/man. This binary gender categorization is limited because it cannot account for intersex, non-binary, or transgender people. In addition, our references contain 6.06\% author of color (first)/author of color(last), 12.12\% white author/author of color, 18.18\% author of color/white author, and 63.64\% white author/white author. Our approach to gender and race categorization is limited in that gender and race are assigned by us based on publicly available information and online media. We look forward to future databases allowing all authors to self-identify race and gender in an appropriately, anonymized, and searchable fashion and new research that enables and supports equitable practices in science.


\subsection*{Data Accessibility.} {The code and data used to produce these results can be found at \url{https://github.com/mjcolebank/CDG_NCSU/}.}

\subsection*{Author Contributions.} {ALC: conceptualization, formal analysis, investigation, methodology, software, validation, visualization, writing—original draft, writing—review and editing;
MJC: conceptualization, formal analysis, investigation, methodology, software, validation, visualization, writing—original draft, writing—review and editing; REU: conceptualization, investigation, and methodology; MSO: conceptualization, investigation, methodology, and editing}

\subsection*{Competing interests.}{The authors declare they have no competing interests.}

\subsection*{Funding.}{The project described was supported by the National Center for Research Resources and the National Center for Advancing Translational Sciences, National Institutes of Health, through Grant \#TL1 TR001415 (MJC), the National Heart Lung Blood Institute, National Institute of Health \#R01HL147590 (MSO),  the National Science Foundation, Division of Mathematical Sciences, National Science Foundation \#1615820 (MSO) and Research Training Group Award \#1246991 (MSO, MJC, REU), and The National GEM Consortium, GEM Graduate Fellowship (ALC). The content is solely the responsibility of the authors and does not necessarily represent the official views of the NIH or NSF. The funders had no role in study design, data collection and analysis, decision to publish, or preparation of the manuscript.}

\subsection*{Acknowledgements.} {Author REU Program include participants: Macie King, Christopher Schell, Matt Sheldon, Mariam Kharbat, and Robert Sternquist who participated in the summer Research Experience for Undergraduates (RTG-REU) summer 2019. We thank Dr. Martin Johnson (Golden Jubile Hospital, Glasgow, Scotland) and Dr. Sudarshan Rajagopal (Duke University) for providing the patient data. }

\bibliographystyle{plain}
\bibliography{refs}

\end{document}